\documentclass[12pt]{article}
\usepackage[OT1]{fontenc}
\usepackage[utf8]{inputenc}
\usepackage[margin=0.9in]{geometry}
\usepackage{graphicx}
\usepackage{algorithm}
\usepackage{algorithmic}
\usepackage{amsmath}
\usepackage{amsfonts}
\usepackage{comment} 
\usepackage{appendix}
\usepackage{subfigure}
\usepackage[usenames,dvipsnames]{xcolor}
\usepackage{lipsum}  
\usepackage{tikz}
\usepackage{amsthm}
\usepackage[affil-it]{authblk}
\usepackage{placeins}

\usepackage{mathrsfs}
\usepackage{amssymb}
\usepackage{mathtools}
\usepackage{multirow}

\title{Multiresolution Approximation of a Bayesian Inverse Problem using Second-Generation Wavelets}
\author[*,1]{Navid Shervani-Tabar}
\affil[1]{Department of Applied and Computational Mathematics and Statistics, University of Notre Dame, Notre Dame, IN 46556, USA}
\affil[*]{Corresponding author: nshervan@nd.edu}
\date{April 2, 2023}

\begin{document}
\maketitle    
% \linenumbers

\begin{abstract}   
    Bayesian approaches are one of the primary methodologies to tackle an inverse problem in high dimensions. Such an inverse problem arises in hydrology to infer the permeability field given flow data in a porous media. It is common practice to decompose the unknown field into some basis and infer the decomposition parameters instead of directly inferring the unknown. Given the multiscale nature of permeability fields, wavelets are a natural choice for parameterizing them. This study uses a Bayesian approach to incorporate the statistical sparsity that characterizes discrete wavelet coefficients. First, we impose a prior distribution incorporating the hierarchical structure of the wavelet coefficient and smoothness of reconstruction via scale-dependent hyperparameters. Then, Sequential Monte Carlo (SMC) method adaptively explores the posterior density on different scales, followed by model selection based on Bayes Factors. Finally, the permeability field is reconstructed from the coefficients using a multiresolution approach based on second-generation wavelets. Here, observations from the pressure sensor grid network are computed via Multilevel Adaptive Wavelet Collocation Method (AWCM). Results highlight the importance of prior modeling on parameter estimation in the inverse problem.
\end{abstract}

\section{Introduction \label{sec:intro}}
Studying fluid flows in porous media is of importance in many applications, including groundwater hydrology and petroleum geology \cite{todd1959ground, aarnes2007modelling}. A system of elliptic Partial Differential Equation (PDE) is generally used to describe physics of the flow in these applications. Primary parameter of interest in these type of problems is the permeability field. However, this information is usually not directly accessible and hence must be inferred from the flow data. This gives rise to an inverse problem.

Inverse problems come to the picture for indirect observations of a quantity of interest. Consider the equation
\begin{equation}
    y_{\textbf{x}}=F(\textbf{x};k)+e_{\textbf{x}}
\end{equation}
where $y_\textbf{x}$ is the observed flow data which is a function of location $\textbf{x}$ and permeability $k$. In a numerical setting, $F$ denotes the numerical solution of the governing PDE and we would like to solve for unknown $k$ given observed data $y_\textbf{x}$. There may be some difficulties with solving these type of problems \cite{dashti2016bayesian}. One problem is that the observed data is prone to perturbation by observational noise $e_\textbf{x}$. In the best case, only information regarding statistical properties of noise are provided and $y_\textbf{x}$ can't be obtained by just a simple subtraction. Another issue is that the system may be underdetermined, meaning that there are more unknowns than equations.

In a Bayesian approach to address inverse problems, observed data $y_\textbf{x}$, unknown $k$, and observational noise $e_\textbf{x}$ are treated as random variables. The noise then can be modeled by considering the solution to the inverse problem as the probability distribution of $k$ given $y_\textbf{x}$, denoted by $k|y_\textbf{x}$. This way, prior modeling can be used to address the underdetermined problem by filling for the missing data.

One way to treat the inverse problem is by decomposing unknown $k$ into sum of bases. This way, coefficients of the expansion of $k$ would be the unknowns. One such decomposition is yield by Fourier transform. Joseph Fourier introduced the idea of expansion or approximation of a periodic function f(x) in terms of trigonometric series. These series, formally known as Fourier series, take advantage of
the orthogonality relationship of the sine and cosine functions.

Beskos et al.~\cite{beskos2015sequential} used Fourier decomposition to infer the unknown parameters of a Bayesian elliptic PDE inversion. However, Fourier series has two major drawbacks. Firstly, it is not intended to represent local information in time. In other words, they capture global behavior of a function. Secondly, although it provides the opportunity to investigate the problem either in spatial domain or frequency domain, it doesn't provide a chance to study both together \cite{debnath2002wavelet}. To overcome these difficulties wavelets were introduced in 1980’s by Grossmann and Marlet \cite{grossmann1984decomposition}. Therefore, wavelets are referred to as a departure from Fourier analysis \cite{hubbard2005world}.

Wavelets are a class of basis functions that are localized in both physical space and wave-number space. Because of these properties, wavelets, unlike the Fourier transform which provides frequency information only, provide both spatial and frequency information. The wavelet family $\psi_{j,n}$, defined as
\begin{equation}
    \psi_{j,n}(x)=2^{-j/2}\psi \big (2^{-j}x- n\big ),
\end{equation}
is the translation and dilation of a mother wavelet, $\psi$, which was primarily constructed using the Fourier transform. Now, the unknown $k$ can be expanded in wavelets orthogonal basis as 

\begin{equation}
    k(x)=\sum_j \sum_n d_{j,n} \psi_{j,n} (x)
\end{equation}
where $d_{j,n}=\langle k, \psi_{j,n} \rangle$ is the wavelet coefficient.

Due to being localized, wavelet coefficients tend to be sparse, which reduces the number of unknowns in our underdetermined problem. This makes them a good choice for parameterizing the unknown in the inverse problems. Ideally, the function $k$ should be reconstructed with only a few basis functions. This means that a few significant coefficients should lead to an accurate representation of the original data. In addition to that, insignificant coefficients are well organized in the form of zero trees. These help to characterize the wavelet transform.

In a Bayesian framework, the wavelet coefficients of unknown $k$ are inferred given a prior distribution of unknowns conditioned on observed data $y_\textbf{x}$. Ellam et al. \cite{ellam2016bayesian} used Daubechies wavelets to decompose the unknown field into wavelet bases. In their work, a uniform prior distribution is imposed on the scaling and wavelet coefficients to implement the Bayesian framework and then scaling and decay rate are placed on these parameters. A uniform prior, however, is a vague prior as it does not provide much information for the Bayesian problem and using it is a missed opportunity to take advantage of the wavelet parameterization characteristics, while such information are available. In other words, this choice of prior does not impose any prior information from the structure of the wavelet coefficients quadtree on the Bayesian model. 

A suitable choice of prior ensures that the forward model is almost surely well defined \cite{beskos2015sequential}. Hence, prior modeling on the wavelet coefficients of unknown $k$ plays an important role in the inverse problem and it should be designed such that it capture the characteristics of the wavelet series representation of the unknown parameter \cite{abramovich1998wavelet}, including the sparseness and the hierarchical relation of the coefficients.

Prior modeling of the wavelet coefficients and their structure has been used in a variety of contexts. 

In the context of denoising, statistical wavelet modeling has been widely used for Bayesian shrinkage of empirical wavelet coefficients. In one of the early works on this topic, Chipman et. al. \cite{chipman1997adaptive} utilized a mixture of Gaussian as the prior distribution on wavelet coefficients and choose hyperparameters of their model based on the empirical wavelet coefficients. Vidakovic \cite{vidakovic1998nonlinear} suggests the use of a heavy tailed prior on the wavelet coefficients population, specifically a student's t distribution. Hyperparameters of this model, which are common for all levels to avoid complexity, were found using a mix of empirical Bayes (for variance) and experimentation (for degree of freedom).

M$\ddot{\text{u}}$ller and Vidakovic \cite{muller1998bayesian} proposed a prior probability model with mixture priors for the wavelet coefficients. The mixture coefficient is set to be level dependent to promote geometrically  increasing  probability  of sparsity  in  wavelet domain.

He and Carin \cite{he2009exploiting} used a spike and slab prior for statistical modeling of wavelet coefficients in the context of compressive sensing. They utilized a hierarchical Bayesian framework to infer the hyperparameter of the prior model via Markov Chain Monte Carlo (MCMC) sampling. In \cite{he2010tree} they further modified their model as a Hadamard product of a Gaussian distribution, which models wavelet coefficients and a Bernoulli distribution, which models the sparseness of the coefficients. 

In a Bayesian framework, inverse problems are explored with sampling methods. Some examples of these methods are Markov Chain Monte Carlo (MCMC) method \cite{hastings1970monte, metropolis1953equation}, Importance Sampling \cite{kahn1950randomI, kahn1950randomII}, and Sequential Monte Carlo (SMC)\cite{doucet2001introduction}. SMC takes advantage of bridging densities $\pi_0, \dots, \pi_M$ to gradually move from prior density $\pi_0=\pi_{pr}$ to posterior density $\pi_M=\pi_{post}$. While working with an expansion in bases, SMC method sequentially moves a prior model of wavelet coefficients to a more complex posterior distribution.

In this work, second generation wavelets, which are generalization of the biorthogonal wavelets, have been used to reconstruct the unknown field. Due to their characteristics including compact support, which leads to a reconstruction with minimal error and increases the probability of capturing events in short instances, second generation wavelets enable us to have a wider selection of priors to implement. Here, we have implemented a spike-and-slab prior, which imposes the prior belief on the sparseness of the scaling and root nodes in the wavelet quadtree structure by setting a mixing weight parameter close to 1 and for all other nodes, by imposing a non-informative value of 0.5. As opposed to \cite{ellam2016bayesian}, No scaling or decay rate is further imposed. Then the posterior for each scale is inferred using sequential Monte Carlo method, which adaptively yields the posterior distribution for each scale. To find the pressure field for any given distribution of wavelet coefficients of the decomposed permeability field, we have taken advantage of multilevel Adaptive Wavelet Collocation Method (AWCM) \cite{vasilyev2005adaptive}, which gets the SMC particles and iteratively solves the forward problem in a physical space on a dynamically adaptive computational grid and interpolates this solution on pressure network grid of desired dimension. 

The rest of the article is organized as following: section \ref{Forward Mathematical model} details the mathematical model for groundwater flow. In section \ref{Multiresolution Analysis} a review of the wavelet basis and multiresolution analysis is presented. In section \ref{AWCM} numerical implementation of the problem is reviewed. In section \ref{Bayesian Approach to Inverse Problems} details of Sequential Monte Carlo method and it's implementation are discussed. In section \ref{Numerical results}, numerical tests have been performed to demonstrate the performance of the algorithm.

\section{Forward Mathematical Model}
\label{Forward Mathematical model}

In an inverse setting, a forward problem describes the relation between model parameters and the observation. In this section, the mathematical model for the forward physical problem is described. Here, the numerical solution of an elliptic partial differential equation, namely steady-state groundwater flow, formulates the forward model. A depiction of flow in porous media from a macro scale point of view is done through the relations

\begin{equation}
\label{eq:elliptic}
\nabla \cdot \boldsymbol{u}(\boldsymbol{x}) = f(\boldsymbol{x}), \quad x \in \mathcal{X},
\end{equation}
\begin{equation}
\label{eq:darcy}
\boldsymbol{u}(\boldsymbol{x}) = -\boldsymbol{k}(\boldsymbol{x}) \nabla p(\boldsymbol{x}), \quad x \in \mathcal{X},
\end{equation}
\begin{equation}
\label{eq:BC}
\boldsymbol{u} \cdot \hat{\boldsymbol{n}} = 0 \text{  on } \partial \mathcal{X},
\end{equation}
Equation \eqref{eq:elliptic} expresses the continuity of mass with $\boldsymbol{u}$ denoting the velocity field and function $f$ representing the source/sink terms,
\begin{equation}
    \label{eq:s/s}
    f(x)=\sum_{i=1}^2 c_i \delta_{x_i} (x)
\end{equation}
where $\delta_{x_i}$ represents a point mass with $x_i$ denoting its position and $c_i$ a sign that indicates whether it is a sink or a source. Here, a source term is positioned at $x_1=[0, 0]$ and a sink is located at $x_2=[1, 1]$. Equation \eqref{eq:darcy} is Darcy flow formulation, where $k$ represents the permeability field and $p$ stands for pressure. Equation \eqref{eq:BC} shows a Neumann type boundary condition which sets a no inflow rule on the boundaries of the flow domain. 

The unknown spatially varying parameter to be inferred and the main unknown of the inverse problem is a permeability field with multiscale characteristics. Together, equations \eqref{eq:elliptic} through \eqref{eq:BC} define an elliptic PDE which solves for pressure $p$ given a permeability field $k$.

% Figure \ref{} illustrates the solution of the forward problem for a uniform permeability field $k=-1/2$ in a $1\times 1$ domain. Given field $k$, the model solves for pressure $p$ and hence, velocity fields in $x$ and $y$ directions are computed using Darcy's law.

The solution of this forward mathematical model would provide the input for the inverse problem. For observed data, this solution is prone to errors, including observational noise. Therefore, an error term defined by realization of uncorrelated Gaussian noise with zero mean is added to the solution of the solver to account for the composite error.
 
\begin{equation*}
    y_{\textbf{x}}=F(\textbf{x};k)+ e_{\textbf{x}}
\end{equation*}
This error term can be modeled using a Gaussian distribution
\begin{equation}
    e_{\textbf{x}}\sim\mathcal{N}(0,\sigma^2)  
\end{equation}

% !TEX root = ../JCP-SCLS.tex
\section{Multiresolution Analysis}
\label{Multiresolution Analysis}

The basic idea behind the wavelet decomposition is to represent a function in terms of basis functions, called wavelets. Wavelets are a class of oscillatory basis functions of finite duration. Unlike sines and cosines that have global support, wavelets are localized in both physical and wave-number space. The need to have both position and frequency information was one of the main motivations that led to the development of wavelets as an alternative to the Fourier transform.

Multiresolution approximation is a sequence of embedded vector spaces for approximating $\textbf{L}^2(\mathbb{R})$ functions using wavelets \cite{mallat2008wavelet}. It computes the approximation of signals at various level of resolution $j$ with orthogonal projections on different spaces  $\{\textbf{V}_j\}_{j\in\mathbb{Z}}$

\begin{equation}
    \{0\}\subset\dots\subset\textbf{V}_{j-1}\subset\textbf{V}_{j}\subset\textbf{V}_{j+1}\subset\dots\subset\textbf{L}^2(\mathbb{R}).
\end{equation}
An orthogonal basis of each space $\textbf{V}_{j}$ is constructed by dilating and translating a single function $\phi$ called scaling function,

\begin{equation}
    \phi_{j,n}(x)=\frac{1}{\sqrt{2^j}}\phi \Big (\frac{x-2^j n}{2^j}\Big ).
\end{equation}
The family $\{\phi_{j,n}\}_{n\in\mathbb{Z}}$ is an orthonormal basis of $\textbf{V}_j$ for all $j\in\mathbb{Z}$. The orthogonal projection of $f$ over $\textbf{V}_j$ is obtained by an expansion in the scaling orthogonal basis

\begin{equation}
    P_{\textbf{V}_j} f=\sum_{n=-\infty}^{+\infty} c_{j,n} \phi_{j,n},
\end{equation}
where the scaling coefficient $c_{j,n}$ is defined using inner product $c_{j,n}=\langle f,\phi_{j,n} \rangle$. Projection of a signal in different spaces $\textbf{V}_j$ gives us the signal in different resolutions $2^j$.

Orthonormal wavelets carry the details necessary to increase the resolution of a signal approximation. $\textbf{W}_j$ is defined as the orthogonal complement of space $\textbf{V}_j$ in $\textbf{V}_{j+1}$,

\begin{equation}
\label{eq:orthComp}
    \textbf{V}_{j+1}=\textbf{V}_{j}\oplus\textbf{W}_{j}.
\end{equation}
An orthonormal basis of $\textbf{W}_j$ is constructed by scaling and translating a function $\psi$, called mother wavelet
\begin{equation}
    \psi_{j,n}(x)=\frac{1}{\sqrt{2^j}}\psi \Big (\frac{x-2^j n}{2^j}\Big ).
\end{equation}
Therefore, for any level of resolution $j$ and suitable choice of $\psi$, the family $\{\psi_{j,n}\}_{n\in\mathbb{Z}}$ is an orthonormal basis of $\textbf{W}_j$. 

From equation \ref{eq:orthComp} it concludes that the orthogonal projection of function $f\in\textbf{L}^2(\mathbb{R})$ on $\textbf{V}_{j+1}$ can be decomposed as the sum of orthogonal projections on $\textbf{V}_{j}$ and $\textbf{W}_{j}$,
\begin{equation}
   P_{\textbf{V}_{j+1}}f= P_{\textbf{V}_{j}}f + P_{\textbf{W}_{j}}f,
\end{equation}
where the complement $P_{\textbf{W}_{j}}f$ is the details of $f$ that appear at the finer level of resolution $j+1$ but that disappear at the coarser level of resolution $j$. The wavelet series representation of $P_{\textbf{W}_{j}}f$ is then
\begin{equation}
    P_{\textbf{W}_j}f=\sum_{n=-\infty}^{+\infty}d_{j,n} \psi_{j,n},
\end{equation}
where the wavelet coefficients $d_{j,n}$ are defined as the inner product $d_{j,n}=\langle f,\psi_{j,n} \rangle$. Therefore, the wavelet series decomposition of $f$ has the form
\begin{equation}
    f(x)=\sum_{n\in \textbf{I}_{\phi}^0}c_{0,n}\phi_{0,n}(x)+\sum_{j=0}^{J-1}\sum_{n\in \textbf{I}_{\psi}^j}d_{j,n}\psi_{j,n}(x)
\end{equation}
where $\phi_{0,n}$ is an orthonormal basis for the reference space $\textbf{V}_0$,  $\textbf{I}_{\phi}^0$ and $\textbf{I}_{\psi}^j$ are index sets associated with functions $\phi_{0,n}$ and $\psi_{j,n}$, and $J$ is the finest level of resolutions within which signal $f$ is approximated.

\subsection{Two-Dimensional Wavelet Bases}
Separable wavelet bases are the best way to design two-dimensional wavelet bases \cite{strang1996wavelets}. The approximation of a two-dimensional signal $f(x_1, x_2)$ at the resolution $2^{j}$ is defined as the orthogonal projection of $f$ on a space $\textbf{V}_j^2$ that is included in $\textbf{L}^2(\mathbb{R})$. For $\textbf{x}=(x_1,x_2)$ and $\textbf{n}=(n_1,n_2)$, 

\begin{equation}
    \bigg\{\phi^2_{j,\textbf{n}}(\textbf{x})= \phi_{j,n_1}(x_1) \phi_{j,n_2}(x_2)=\frac{1}{2^j}\phi \Big( \frac{x_1-2^j n_1}{2^j} \Big) \phi \Big ( \frac{x_2-2^j n_2}{2^j} \Big )\bigg\}_{\textbf{n}\in\mathbb{Z}^2}
\end{equation}
is an orthonormal basis of $\textbf{V}_j^2$. This basis is obtained by scaling by $2^j$ the two-dimensional separable scaling function $\phi^2(\textbf{x})= \phi(x_1) \phi(x_2)$ and translating it on a two-dimensional square grid with intervals $2^{-j}$. 

A separable wavelet orthonormal basis of $\textbf{L}^2(\mathbb{R}^2)$\ is constructed with separable products of a scaling function $\phi$ and a wavelet $\psi$. Let $\textbf{W}_j$ be the detail space equal to the orthogonal complement of the lower-resolution approximation space $\textbf{V}_j^2$ in $\textbf{V}_{j+1}^2$, 

\begin{equation}
    \textbf{V}^2_{j+1}=\textbf{V}^2_{j}\oplus\textbf{W}^2_{j}.
\end{equation}
Decomposing $\textbf{V}_{j+1}$ we can show that

\begin{equation}
    \textbf{W}^2_j=(\textbf{V}_j \otimes \textbf{W}_j)\oplus(\textbf{W}_j \otimes \textbf{V}_j)\oplus(\textbf{W}_j \otimes \textbf{W}_j).
\end{equation}
Since $\{\phi_{j,m}\}_{m\in\mathbb{Z}}$ and $\{\psi_{j,m}\}_{m\in\mathbb{Z}}$ are orthonormal bases of $\textbf{V}_j$ and $\textbf{W}_j$, the wavelet family 

\begin{equation}
    \bigg \{ \psi^1_{j,\textbf{n}}, \psi^2_{j,\textbf{n}}, \psi^3_{j,\textbf{n}}\bigg \}_{\textbf{n}\in\mathbb{Z}^2}
\end{equation}
where 
\begin{align}
    \psi^1(\textbf{x}) &= \phi(x_1) \psi(x_2),\\   \psi^2(\textbf{x}) &= \psi(x_1) \phi(x_2),\\   \psi^3(\textbf{x}) &= \psi(x_1) \psi(x_2),
\end{align}
is an orthonormal basis of $\textbf{W}_j^2$. $\psi^1$, $\psi^2$, and $\psi^3$ show wavelets in horizontal, vertical, and diagonal directions, respectively. We denote the family $\psi_{j,n}^\mu$ by

\begin{equation}
    \psi^\mu_{j,\textbf{n}}(\textbf{x})=\frac{1}{2^j}\psi^\mu \Big (\frac{x_1-2^j n_1}{2^j}, \frac{x_2-2^j n_2}{2^j}\Big ),
\end{equation}
for $1\leq \mu \leq 2^D-1$ with $D$ denoting the dimension. Therefore, the wavelet decomposition of two-dimensional signal $f$ has the form
\begin{equation}
\label{eq:2ddecomp}
    f(\textbf{x})=\sum_{\textbf{n}\in \textbf{I}_{\phi}^0}c_{0,\textbf{n}}\phi_{0,\textbf{n}}(\textbf{x})+\sum_{j=0}^{J-1}\sum_{\mu=1}^{2^D-1}\sum_{\textbf{n}\in \textbf{I}_{\psi}^j}d^\mu_{j,\textbf{n}}\psi^\mu_{j,\textbf{n}}(\textbf{x})
\end{equation}
In this equation, coefficient $c_{0,\textbf{n}}$ gives an approximation of two-dimensional signal $f$ at the reference space $\textbf{V}^2_0$. Large amplitude coefficients in $d^1$, $d^2$, and $d^3$ correspond to details in vertical (vertical high frequencies), horizontal, and diagonal directions (high frequencies in both directions), respectively.

\subsection{Second-Generation Wavelets}

Second-generation wavelets are a generalization of biorthogonal wavelets, which are more easily applied to functions defined on domains more general than $\textbf{R}^n$. Their major benefit comes from their construction in the spatial domain. This helps second-generation wavelets to be tailor maid for complex domains and irregular sampling \cite{vasilyev2000second}.

Family of wavelets $\psi_{j,k}$ are classically constructed by translation and dilation of one mother wavelet $\psi$. The lack of translation and dilation invariance, however, does not allow for classical way in such more general cases as boundaries and irregular sampling \cite{sweldens1998lifting}. A better way to build wavelets in these cases comes from abandoning the translation/dilation approach. The wavelet resulted by the departure from classical translation/dilation method is referred to as second-generation wavelets. Rather than the Fourier transform, main tools for constructing second-generation wavelets include interpolating wavelet transform and lifting schemes

Interpolating wavelet transform \cite{donoho1992interpolating, harten1994adaptive} can best be explained by an example on transforming data to different levels of resolution. Consider the dyadic grid

$$ \mathcal{G}^j=\{x_{j,n}\in \mathcal{R}:x_{j,n}=2^{-j}n, n\in Z\}, j\in Z$$
where $x$ and $j$ denote grid points and level of resolution, respectively. Given discrete data $f(x_{j,n})$ we can project the data on a finer grid $\mathcal{G}^{j+1}$ by predicting the value of data sequence on all dyadic points in between. This prediction can be achieved by interpolation. Here, we take advantage of polynomial interpolation of order $2N-1$ using $2N$ closest neighbour points. Using this method, the signal can be approximated on finer grid of level $j+1$ as 

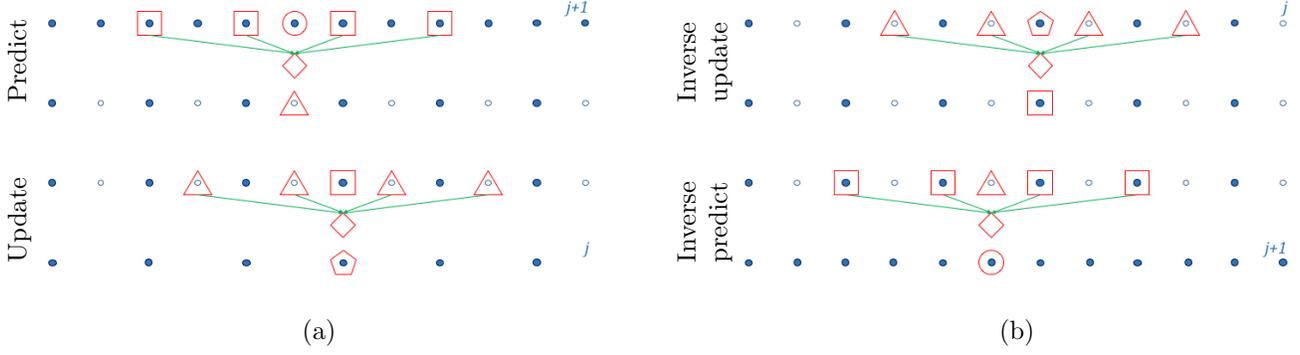
\begin{figure}[!t] 
\setlength{\unitlength}{0.01\textwidth} 
\begin{picture}(100,26)
\put(0.5,0){\includegraphics[width=0.48\textwidth]{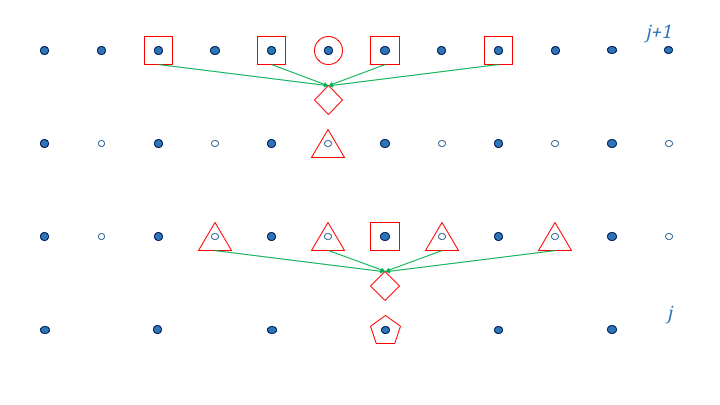}}
\put(55.0,0){\includegraphics[width=0.48\textwidth]{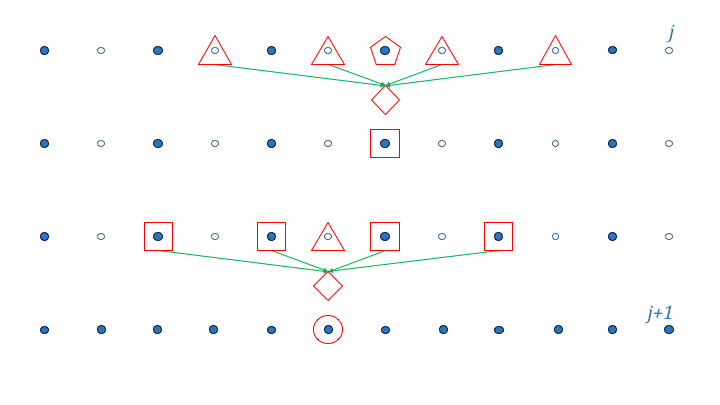}}
\put(23,-1){\footnotesize (a)}
\put(77.5,-1){\footnotesize (b)}
\put(0,5){\rotatebox{90}{\footnotesize Update}}
\put(0,17.5){\rotatebox{90}{\footnotesize Predict}}
\put(52,5){\rotatebox{90}{\footnotesize {\begin{tabular}[c]{@{}c@{}}Inverse\\ predict\end{tabular}}}}
\put(52,17.5){\rotatebox{90}{\footnotesize {\begin{tabular}[c]{@{}c@{}}Inverse\\ update\end{tabular}}}}
\end{picture}
\caption{Forward (a) and inverse (b) wavelet transform using second-generation wavelets. \label{fig:fwd_inv} 
 }
\end{figure}

\begin{equation}
    f^j(x_{j+1,2n+1})=\sum_{l=-N+1}^N \gamma_{n,l}^jf(x_{j,n+l})
    \label{eq:interp}
\end{equation}
where $\gamma$ denotes the interpolation weight. Assuming that the function $f^j (x_{j, n})$ belongs to the space $\textbf{V}_j$, the interpolated function at level $j+1$ can be shown by $f^{j} (x_{j+1, n})$. Since $x_{j,n} = x_{j+1,2n}$ we can show that for the even points, $f^j(x_{j+1,2n})=f^{j+1}(x_{j+1,2n})$. However, the prediction at odd points might not be accurate and hence, $f^j(x_{j+1,2n+1})\neq f^{j+1}(x_{j+1,2n+1})$. So, the details necessary for accurate reconstruction of $f^j(x)$ on grid $\mathcal{G}^{j+1}$ are the differences between $f^j(x_{j+1,2n+1})$ and $f^{j+1}(x_{j+1,2n+1})$. If we call half the difference $f^{j+1}(x_{j+1,2n+1})-f^j(x_{j+1,2n+1})$ a wavelet coefficient $d_{j,n}$, then we see that $f^{j+1}(x) = f^j (x) + d_j (x)$.

The wavelet basis constructed using interpolating scaling functions, however, has a non-zero mean and hence, does not provide a Reisz basis for $L^2$.
%and the dual wavelets are Dirac $\delta$-functions which do not belong to $L^2$. 
Also, from a signal processing point of view, interpolating scaling functions have a constant as low-pass filter and hence, are unable to well separate the scales. Consequently, this wavelet transform would introduce numerical instabilities and aliasing \cite{vasilyev2000second}. To overcome these difficulties, lifting scheme is used to enhance the properties of wavelets.

As stated earlier, the next tool needed for constructing second-generation wavelets is lifting scheme \cite{sweldens1998lifting}. Lifting scheme starts from a simple multiresolution analysis and by adding linear combination of scaling function to the wavelet of same level of resolution builds $a$ $priori$ defined properties into it. 

In summary, the two steps for forward wavelet transform using second-generation wavelet can be formulated as
\begin{align}
    &\text{Predict Stage:}\qquad d_{j,n}=\frac{1}{2}\Big(c_{j+1,2n+1}-\sum_l \gamma^j_{k,l}c_{j+1, 2n+2l}\Big) \label{eq:fwdPrd}\\
    &\text{Update Stage:}\qquad c_{j,n}=c_{j+1,2n}+\sum_l \tilde{\gamma}^j_{k,l}d_{j, n+l} \label{eq:fwdUpd}
\end{align}

\begin{algorithm}[H]
	\caption{Forward wavelet transform using second-generation wavelets. \label{alg:fwdWlt}}
	\begin{algorithmic}
	\FOR{$j=J:j_{min}+1$}
        \STATE \textbf{\textcolor{ForestGreen}{predict}} odd points $x_{j,2k+1}$ using $2N$ closest even points $f^j(x_{j,2k})$ using Eq. \ref{eq:interp}.
        \STATE \textbf{calculate} wavelet coefficients $d_{j-1,k}$ using Eq. \ref{eq:fwdPrd}.
        \STATE \textbf{assign} wavelet coefficients $d_{j-1,k}$ to odd points $x_{j,2k+1}$.
        \STATE \textbf{interpolate} even points $x_{j,2k}$ using $2N$ closest odd points $x_{j,2k+1}$.
        \STATE \textbf{\textcolor{NavyBlue}{update}} even points $x_{j,2k}$ using Eq. \ref{eq:fwdUpd}.
        \STATE \textbf{save} values on even points $x_{j,2k}$ as scaling coefficients $c_{j-1,k}$.
    \ENDFOR
	\end{algorithmic}
\end{algorithm}

The output of the update stage is a wavelet with zero mean, which results in an accurate transform.

Next, the inverse wavelet transform is performed using the same stages and operations, but in reversed order. In other words, first, even points at level $j+1$ are calculated at the inverse update stage. Then, the odd points are found after inverse predict stage. These relations are shown using equations \ref{eq:invUpd} and \ref{eq:invPrd}. Figure \ref{fig:fwd_inv} summarizes forward and inverse transforms.

\begin{align}
    &\text{Inverse Update Stage:}\qquad c_{j+1,2k}=c_{j,k}-\sum_l \tilde{\gamma}^j_{k,l}d_{j,k+l} \label{eq:invUpd}\\
    &\text{Inverse Predict Stage:}\qquad c_{j+1,2k+1}=2d_{j,k}+\sum_l \tilde{\gamma}^j_{k,l}c_{j+1,2k+2l}\label{eq:invPrd}
\end{align}

\begin{algorithm}[H]
	\caption{Inverse wavelet transform using second-generation wavelets. \label{alg:invWlt}}
	\begin{algorithmic}
	\FOR{$j=j_{min}:J-1$}
        \STATE \textbf{assign} wavelet coefficients $d_{j,k}$ to odd points $x_{j+1,2k+1}$.
        \STATE \textbf{assign} scaling coefficients $c_{j,k}$ to even points $x_{j+1,2k}$.
        \STATE \textbf{interpolate} even points $x_{j+1,2k}$ using $2N$ closest odd points $x_{j+1,2k+1}$.
        \STATE \textbf{\textcolor{NavyBlue}{inverse update}} $f^{j+1}(x_{j+1,2k})$ using Eq. \ref{eq:invUpd}.
        \STATE \textbf{interpolate} odd points $x_{j+1,2k+1}$ using $2N$ closest even points $x_{j+1,2k}$.
        \STATE \textbf{\textcolor{ForestGreen}{inverse predict}} $f^{j+1}(x_{j+1,2k+1})$ using Eq. \ref{eq:invPrd}.
    \ENDFOR
	\end{algorithmic}
\end{algorithm}

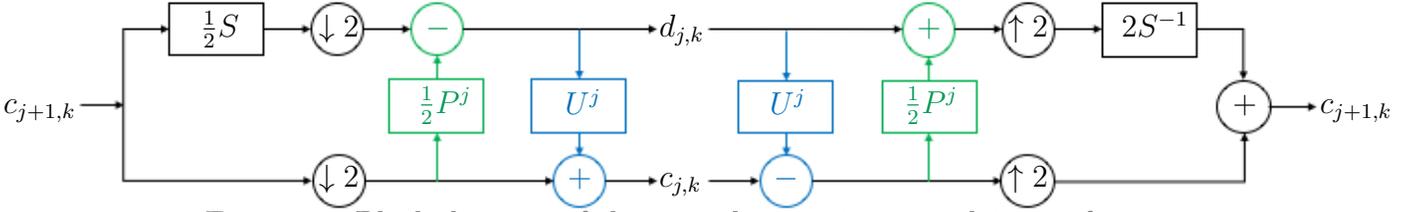
\begin{figure}[!t] 
\setlength{\unitlength}{0.01\textwidth} 
\begin{picture}(100,17)
\put(0.1,-4.3){\includegraphics[width=1.01\textwidth]{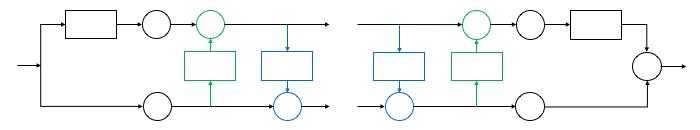}}
% top row
\put(12,10.35){$\frac{1}{2}S$}
\put(21.25,10.35){$\downarrow 2$}
\put(29.6,10.35){\textcolor{ForestGreen}{$-$}}
\put(47.85,10.35){$d_{j,k}$}
\put(68,10.35){\textcolor{ForestGreen}{$+$}}
\put(75.1,10.35){$\uparrow 2$}
\put(84,10.35){$2S^{-1}$}
% middle row
\put(-3.4,4.4){$c_{j+1,k}$}
\put(29,4.4){\textcolor{ForestGreen}{$\frac{1}{2}P^j$}}
\put(40.5,4.3){\textcolor{NavyBlue}{$U^j$}}
\put(56.5,4.3){\textcolor{NavyBlue}{$U^j$}}
\put(67,4.4){\textcolor{ForestGreen}{$\frac{1}{2}P^j$}}
\put(92.7,4.4){$+$}
\put(99.5,4.4){$c_{j+1,k}$}
% buttom row
\put(21.25,-1.4){$\downarrow 2$}
\put(40.7,-1.4){\textcolor{NavyBlue}{$+$}}
\put(47.85,-1.4){$c_{j,k}$}
\put(56.8,-1.4){\textcolor{NavyBlue}{$-$}}
\put(75.1,-1.4){$\uparrow 2$}
\end{picture}
\caption{ Block diagram of the second generation wavelet transform. \label{fig:fwd_inv} 
 }
\end{figure}

\FloatBarrier

% !TEX root = ../JCP-SCLS.tex
\section{Numerical Implementation}
\label{AWCM}

\subsection{Adaptive Wavelet Collocation Method}
In this study, a general AMR-type numerical method for solving partial differential equations \cite{vasilyev2000second} based on arbitrary order bi-orthogonal second-generation wavelets \cite{sweldens1998lifting} has been used to solve the Darcy flow equation. For the reader's convenience the Adaptive Wavelet Collocation Method (AWCM) is briefly reviewed next. For the detailed description of the method and its parallel implementation the reader is referred to Refs.~\cite{vasilyev2000second, vasilyev2003solving, vasilyev2005adaptive, nejadmalayeri2015parallel}. 

A multiresolution wavelet representation of a field $u(\textbf{x})$ can be formally written as
\begin{equation}
u(\textbf{x})=\sum_{\textbf{k}\in \textbf{I}_\phi^0}\bar{u}_\textbf{k}^0\phi_\textbf{k}^0(\textbf{x})+\sum_{j=0}^{+\infty}\sum_{\mu=1}^{2^d-1}\sum_{\textbf{k}\in \textbf{I}_\psi^{\mu , j}}\tilde{u}_\textbf{k}^{\mu , j}\psi_\textbf{k}^{\mu , j}(\textbf{x}),
\label{eqwave21}
\end{equation}
where  $\textbf{I}_\phi^0$ and
$\textbf{I}_\psi^{\mu , j}$ are $d$-dimensional index sets associated with scaling
functions $\phi_{\mathbf l}^{0}$ and wavelets $\psi_{\mathbf k}^{\mu, j}$, respectively, $\bar{u}_\textbf{k}^0$ and $\tilde{u}_\textbf{k}^{\mu , j}$ represent scaling and wavelet coefficients, while  $\phi_\textbf{k}^0(\textbf{x})$ and $\psi_\textbf{k}^{\mu , j}(\textbf{x})$ respectively express scaling functions at the coarsest level of resolution and wavelet basis functions of different family, $\mu$, and levels of resolution, $j$.

Multiresolution decomposition \eqref{eqwave21} provides a natural platform to construct adaptive numerical methods based on the analysis of the wavelet coefficients, $\tilde{u}_\textbf{k}^{\mu, j}$, which, in general, have small values except for the regions close to the large gradients.
Formally, the field $u(\textbf{x})$ can be decomposed into two sub-fields
\begin{equation}
u(\textbf{x})=u_\geqslant(\textbf{x})+u_<(\textbf{x})
\label{eqwave5}
\end{equation} 
defined by
\begin{equation}
u_{\geqslant}(\textbf{x})=\sum_{\textbf{k}\in \textbf{I}_\phi^0}\bar{u}_\textbf{k}^0\phi_\textbf{k}^0(\textbf{x})+\sum_{j=0}^{+\infty}\sum_{\mu=1}^{2^d-1}\sum_{\substack{\textbf{k}\in \textbf{I}_\psi^{\mu , j} \\ |\tilde{u}_\textbf{k}^{\mu , j}|\geqslant\epsilon\|u\|}}\tilde{u}_\textbf{k}^{\mu , j}\psi_\textbf{k}^{\mu , j}(\textbf{x})
\label{eqwave3}
\end{equation}
and
\begin{equation}
u_<(\textbf{x})=\sum_{j=0}^{+\infty}\sum_{\mu=1}^{2^d-1}\sum_{\substack{\textbf{k}\in \textbf{I}_\psi^{\mu , j} \\ |\tilde{u}_\textbf{k}^{\mu , j}|<\epsilon\|u\|}}\tilde{u}_\textbf{k}^{\mu , j}\psi_\textbf{k}^{\mu , j}(\textbf{x}) ,
\label{eqwave4}
\end{equation}
where $\epsilon>0 $  is the non-dimensional (relative) thresholding parameter defining the decomposition  and $\left\lVert  u \right\rVert$ is the (absolute) dimensional  scale.  This characteristic amplitude scale is often taken as either the $L_2$- or $L_\infty$-norm of the field $u$ or its derived quantity of interest, e.g., fluctuating component.
 As a result of this split, the compression can be achieved by keeping only the wavelets with coefficients greater than $\epsilon \|u\|$. In other words, high resolution computations are only performed where it is necessary and a significant reduction in the number of required wavelets can be achieved. The AWCM takes advantage of the wavelet compression properties \eqref{eqwave3} and one-to-one correspondence between wavelets and the corresponding grid points on a multi-level computational mesh. As a result the AWCM has the ability to identify, isolate, and track localized, dynamically dominant flow structures, such as interfaces, on adaptive computational mesh while {\em a priori} controlling the accuracy of the solution at the desired level $O(\epsilon)$. This property has distinguished wavelet techniques from conventional methods \cite{schneider2010wavelet}.  

\subsection{Multilevel AWCM}
In this work, an extension of AWCM has been used for the solution of the elliptic problem. In this section, we briefly discuss the method. For a detailed discusion, reader is refered to \cite{vasilyev2005adaptive}. A linear elliptic PDE may be written in the general form
\begin{equation}
    \mathscr{L}\textbf{u}=\textbf{f},
\end{equation}
where $\mathscr{L}$ is a linear elliptic operator (including boundary conditions), and $f$ is a source term. The goal is determining u to within a specified residual tolerance $\|\textbf{u}-\textbf{f}\|_p<\epsilon$ given $\mathscr{L}$ and $f$. Using wavelet-based adaptive mesh refinement assumes the knowledge of the solution in the highest level of resolution. However, this coincides with introduction of computational overhead, especially for highly localized solution. For this purpose the elliptic multilevel wavelet collocation solver is used. In this algorithm, the PDE is iteratively solved starting from the coarsest resolution and the grid of collocation points is continuously refined to resolve the local structures that appear in the solution.

Using AWCM, the adaptive computational grid $\mathscr{G}_{\geq}=\mathscr{G}^j_{\geq}$ is constructed as a set of nested adaptive computational grids $\mathscr{G} ^j_{\geq} \subset \mathscr{G} _{\geq}$, such that $\mathscr{G}^j_{\geq}\subset\mathscr{G}^{j+1}_{\geq}$, for any  $j<J-1$, where $J$ is the finest level of resolution. This nested grid structure provides a framework that allows the approximation from coarser levels of resolution to be used to improve the approximation at the finest level. 

If we use coarse grid correction to solve the coarse grid system, the resulting recursive algorithmic scheme is called a multilevel V-cycle. The multilevel iterative algorithm uses lower-order wavelet differentiation for the approximate solver. Moreover, wavelet interpolation and projection are used respectively for prolongation (injection) and restriction operators. Algorithm \ref{alg:multilevel} summarizes the multilevel AWCM.

\begin{algorithm}[H]
	\caption{Multilevel AWCM \label{alg:multilevel}}
	\begin{algorithmic}  
        \STATE \textbf{initial guess} ($m=0$) : $\textbf{u}^m_\textbf{k}$ and $\mathscr{G}^m_{\geq}$.
        \WHILE{$(m=0)$ \textbf{or} $(m\geq 1$ \textbf{and} $\mathscr{G}^m_{\geq}\neq \mathscr{G}^{m-1}_{\geq}$ \textbf{or} $\|\textbf{u}_\textbf{k}^m-\textbf{u}_\textbf{k}^{m-1}\|_\infty)$}
            \STATE \textbf{perform} forward wavelet transform for each component of $\textbf{u}_\textbf{k}^m$
            \STATE \textbf{construct} $\mathscr{G}^{m-1}_{\geq}$ from significant wavelets $\|d_\textbf{l}^{\mu,j}\|\geq\epsilon$, their neighbours, ghost points, etc.
            \IF {$\mathscr{G}^{m+1}_{\geq}\neq \mathscr{G}^m_{\geq}$}
                \STATE \textbf{interpolate} $\textbf{u}_\textbf{k}^m$ to $\mathscr{G}^{m+1}$.
            \ENDIF
            \WHILE{$\|f^{J}-\mathscr{L}\textbf{u}^J_{\geq}\|_{\infty}>\delta_{\epsilon}$}
                \STATE \textbf{calculate} residual $r^{J}=f^{J}-\mathscr{L}\textbf{u}^J_{\geq}$
                \FOR {$j=J:j_{min}+1$}
                    \STATE \textbf{solve} $\mathscr{L}\textbf{v}^j=\textbf{r}^j$
                    \STATE \textbf{restrict} residual $r^j$ to coarser grid $\mathscr{G}^{j-1}$ using wavelet projection.
                \ENDFOR
                \STATE \textbf{solve} $\mathscr{L}\textbf{v}^{j_{min}}=\textbf{r}^{j_{min}}$
                \FOR {$j=j_{min}+1:J$}
                    \STATE \textbf{prolong} coarse grid solution $v^{j-1}$ to finer grid $\mathscr{G}^{j}$ using wavelet interpolation.
                    \STATE \textbf{solve} $\mathscr{L}\textbf{v}^j=\textbf{r}^j$
                \ENDFOR
                \STATE $\textbf{u}^J_{\geq} = \textbf{u}^J_{\geq} + \textbf{v}^J$
            \ENDWHILE

            \STATE $m=m+1$
        \ENDWHILE
	\end{algorithmic}
\end{algorithm}

\begin{figure}[h]
	\centering
	\includegraphics[width=9cm]{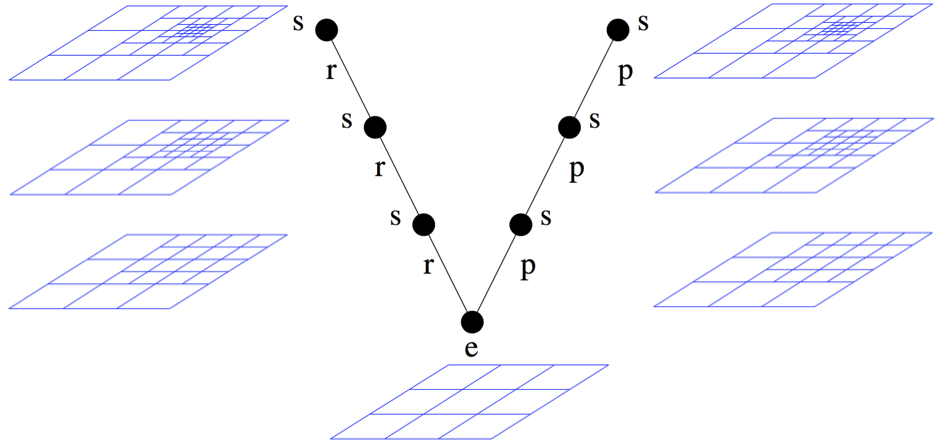}
	\caption{Schematic view of the multilevel V-cycle}
\end{figure}

% !TEX root = ../JCP-SCLS.tex
\section{Bayesian Approach to Inverse Problems}
\label{Bayesian Approach to Inverse Problems}

In this section, our Bayesian approach to answer the inverse problem is described. In the Bayesian framework, we treat the unknown coefficients of the decomposed log-permeability field as random variables and we infer them from the observation data provided from multilevel AWCM solver. 

Using the Bayes rule we have
\begin{equation}
    \pi_{post}=\frac{1}{Z}{\mathcal L}(\textbf{d};\textbf{y})\pi_{pr}(\textbf{d})
    \label{BayesRule}
\end{equation}
where $\mathcal L$ denotes likelihood, $\pi_{pr}$ stands for prior distribution, $\pi_{post}$ shows posterior distribution, and $Z$ is the normalization constant. 

To define the Bayesian model, first prior probability distribution of wavelet coefficients is explored in section \ref{subsec:prior}. Then, in section \ref{subsec:Likelihood} likelihood is derived by marginalizing the observation over noise precision. Finally, normalization constant is used to perform the task of model selection, as illustrated in section \ref{subsec:ModelSelection}.

\subsection{Prior modeling of wavelet coefficients}
\label{subsec:prior}
A prior distribution imposes our belief of the quantity of interest before any observation is provided. In an underdetermined inverse problem, this information helps with filling for missing data. Therefore, prior modeling of the wavelet coefficients can help with finding a faster and more accurate solution. 

When the unknown is parameterized in some basis, prior distribution should include information regarding the coefficients of this parameterization, which are now the unknowns of the problem. The knowledge of the structure of the wavelet coefficients has previously been used in such applications as image compression \cite{said1996new} and compressive sensing \cite{he2009exploiting}. A two-dimensional wavelet transform forges a quadtree of wavelet coefficients, which is consist of several subbands that are grouped into different levels of details, using a series of low-pass and high-pass filters. Each subband contains coefficients corresponding to wavelets in either horizontal, vertical, or diagonal directions with hierarchical relation between subbands of different levels of resolution. 

The hierarchical relation means that in a wavelet quadtree starting from the root nodes all the way to the nodes before the leaf nodes each wavelet coefficient in a coarser scale has 4 children in the next finer level. Leaf nodes have the highest possible resolution level and hence have no children. Wavelet coefficients whose parent is zero tend to have zero value. These zeros then propagate through all finer scales. In most cases, hence, negligible coefficients form a cluster which is referred to as a zero tree.

Due to these characteristics of a quadtree of wavelet coefficients, subbands become more and more sparse as they approach the finer scales. Hence, most of the coefficients in the finer scales are expected to have a value close to zero. This means that a few coefficients with large value contain most of the information from the original data. Therefore, one may reconstruct the field by only using the large coefficients and yet get an accurate representation.

In order to introduce these properties in a statistical manner, we use a spike-and-slab prior

%This spike and slab prior is consist of a positive prior probability mass at zero which leads to a posteriori thresholding and generally to a posteriori shrinkage on the coefficients.

%decreasing prior probabilities for

\begin{equation}
    \label{eq:prior}
    w_{i} \sim \gamma_{s}^{s} \mathcal{N}\left(0, r^{s} \alpha_{s}^{-1}\right)+\left(1-\gamma_{s}^{s}\right) \delta_{0}
\end{equation}
where $\delta_0$ is mass point at zero. Hyperparameters $\gamma_s$, $r$, and $\alpha_s$ denote the mixing factor, scaling factor, and the precision parameter, respectively, and $s$ stands for the level of details. This prior imposes the belief that with probability $\gamma_s^s$ the coefficients have a Gaussian distribution and with probability $1-\gamma_s^s$ it is set to zero which effectively incorporates our knowledge about the sparsity of the wavelet coefficients into the problem.

Factor $r$ explicitly decreases the prior probability of large coefficients in the higher levels of resolution. In addition, mixing coefficients introduces a geometrically decreasing probability for non-negligible coefficients. Combined, these factors introduce a rate of decay on the wavelet coefficients to ensure smoothness of the reconstructed function. As oppose to \cite{ellam2016bayesian} which scales the posterior wavelet coefficients, this implements the smoothness property in the prior model.

In order to represent the quadtree structure of the wavelet coefficients of log-permeability field, the hyperparameters are level-dependent and set according to each scale. For instance, the scaling coefficients ($s=0$) tend to have non-zero values. As a result, a unit mixing coefficient is desired. Note that this is always true due to the power $s$ of the mixing factor. For the root node, most coefficients are non-zero. So we choose a value close to one. For all other levels of resolution, we impose a non-informative value of $0.5$. This indicates that with a probability $0.5$, wavelet coefficients are drawn from a Gaussian distribution and with the same probability, coefficients are set to zero.

Note that, with this choice of prior model, zero valued coefficients propagate through all the finer scales as each such coefficient has four zero valued children in the next finer scale and the rest of the coefficients in that finer scale are set to zero with a probability of $0.5$.

\subsection{Likelihood}
\label{subsec:Likelihood}

% Considering each observation as a random variable conditioned on $\textbf{d}$ we can derive following likelihood function

Using the prior distribution defined in section \ref{subsec:prior}, we would like to calculate the posterior distribution of scaling and wavelet coefficients given the observation. According to the Bayes rule
\begin{align}
    P_{Post}(w|d)\propto P(d|w)P_{Pr}(w)
\end{align}
The goal is to find the likelihood $\mathcal{L}(w ; d)$, which yields a posterior belief of the coefficients of the unknown field by updating prior distribution using observed data. Considering the observational noise, we can model our observation as
\begin{equation}
    d|w, \alpha_e \sim \mathcal{N}(F(w), \alpha_e^{-1})\\
\end{equation}
with $\alpha_e$ denoting the precision of the observational noise. This hyperparameter can be modeled using
\begin{equation}
    \alpha_e | a_0, b_0 \sim \mathcal{G}amma (a_0, b_0)
\end{equation}
where following \cite{gelman2006prior} hyperparameters $a_{0}$ and $b_{0}$ are set to have a non-informative value of 0.001. Therefore, the conditional density of $d$ given $w$ is derived by marginalizing the observation over $\alpha_e$
\begin{equation}
    P(d\mid w) = \int^\infty_0 P(d\mid w, \alpha_e) P(\alpha_e\mid a_0, b_0)d\alpha_e
\end{equation}
Hence, considering each observation as a random variable conditioned on $\textbf{d}$ we can derive following likelihood function

\begin{equation}
   {\mathcal L}(\textbf{w}; \textbf{d})=\big(b_0+\frac{1}{2}\| \textbf{d}-\textbf{F(w)}\|^2_2\big)^{(a_0+n_d/2)}
\end{equation}

To find the posterior distribution, in addition to this likelihood and the prior defined using equation \ref{eq:prior}, we still need to find the normalization constant $Z$. We will use Sequential Monte Carlo method (SMC) for numerical evaluation of this constant. 

\subsection{Performing SMC}

Analytical solution of the posterior defined by the equation~\ref{BayesRule} using the prior and likelihood described in \ref{subsec:Likelihood} and \ref{subsec:prior} is not feasible. Here, we describe the SMC approach used for exploring the posterior density. To facilitate the procedure, auxiliary equations

\begin{equation}
    \pi_{t}(\textbf{w})=\frac{1}{Z_t}{\mathcal L}^{\gamma_t}(\textbf{w};\textbf{d})\pi_{pr}(\textbf{w})
\end{equation}
have been used, where $t=0, \dots M$. These M+1 bridging densities $\pi_0, \dots, \pi_M$ gradually move from prior density to posterior density. This provides a smooth transition between prior and posterior densities. At each increment $t$ we update the importance weights using

\begin{equation}
    \omega_t=\omega_{t-1}\frac{\pi_{t}(\textbf{w}_{t-1})}{\pi_{t-1}(\textbf{w}_{t-1})}
\end{equation}

\subsection{Random walk Metropolis-Hastings}

To rejuvenate the particles, we perform a random walk Metropolis-Hastings algorithm with target distribution $\pi_t$. To do this, we define the random variable
\begin{equation}
    U\sim{\mathcal U}(0, 1)
\end{equation}
Then we draw rejuvenated particle

\begin{equation}
    w'\sim{\mathcal N}(w_t, \sigma_t^2 \textbf{I})
\end{equation}
If $U$ is less than $\min(1, \frac{\pi_t(w')}{\pi_t(w_t)})$, we rejuvenate the particle. Note that, in order to keep the acceptance rate $\alpha_t$ of the random walk proposal around $20\%-30\%$, we adjust the step size of the random walk as following:

\begin{equation}
\sigma_{t+1}= \begin{cases}2 \sigma_t, & \text { if } \alpha_t>0.30 \\ \frac{1}{2} \sigma_t, & \text { if } \alpha_t>0.15 \\ \sigma_t, & \text { otherwise }\end{cases}
\end{equation}

\subsection{Resampling}
\label{subsec:Resampling}

At each iteration, particles with lower weights are less likely to be the solution. To ensure that the computational resources are not dedicated to unlikely particles and that more likely particles have more resources dedicated to them, particles need to be redistributed. A good measure of degeneracy in particles is defined using Effective Sample Size threshold (ESS)

\begin{equation}
\text{ESS}\coloneqq\frac{1}{\sum_{i=1}^N(\bar{\omega}_t^{(i)})^2}
\end{equation}
where $\bar{\omega}$ shows normalized weights. Here, the threshold is set to be $0.95N$. If ESS falls below this value, particles will be resampled using multinomial resampling. 

\subsection{Model selection}
\label{subsec:ModelSelection}

In this problem, we adaptively infer the log permeability field at each scale, i.e. solution at each scale is used as a prior to infer solution at the next scale. In order to choose the scale at which we terminate the algorithm, a model selection criterion based on Bayes Factor is used. Bayes Factor is defined as

\begin{equation}
    \text{BF}_{s,s-1} \coloneqq \frac{Z_s}{Z_{s-1}} = \frac{\pi(\textbf{d}|\mathcal{M}_s)}{\pi(\textbf{d}|\mathcal{M}_{s-1})}
\end{equation}
where
\begin{equation}
  \frac{Z_s}{Z_{0}} =\frac{1}{N}\sum_{i=1}^{N}\omega^i_s
\end{equation}

Bayes factor provides us means of comparing different models, which here are the log-permeability fields at different scales. To maintain this ratio throughout the sampling and after each resampling, we reset importance weights to the average of the importance weights at that increment. Furthermore, importance weights at each scale $s$ are initialized with the average of the importance weights before resampling at the last increment of the previous scale $s-1$.  
%%% Local Variables:
%%% mode: latex
%%% TeX-master: "../JCP-SCLS"
%%% End:
% !TEX root = ../JCP-SCLS.tex
\section{Numerical results}
\label{Numerical results}
This section includes the results of the adaptive scale determination algorithm for various benchmark tests. In the prior, the value of mixing factor $\gamma_0$ is set to $1$ for the scaling coefficients. This is due to the fact that scaling coefficients usually have non-zero values. It is noteworthy that this would have naturally yield due to the zero power in the mixing weight for $s=0$. For the root node, this value is close to $1$. We have set this to value $\gamma_1=0.85$. For all other scales, if the parent coefficients are non-zero, before applying the power $s$, we impose same probability for the coefficients to be zero or non-zero by setting $\gamma_s$ for $s>1$ to be $0.5$. Then, power $s$ introduces a geometrically decreasing probability for significant coefficients. Lastly, if the parent coefficients are zero, we set the child coefficients to be zero. The values of variance $\alpha_s^{-1}$ and scaling factor $r$ are set to $0.7$ and $0.5$, respectively, for all levels of resolution.

To eliminate the periodicity resulting from the use of periodic wavelets, we apply inverse wavelet transform on a $64\times64$ grid in a $[0,2]\times[0,2]$ domain and discarded the extra points to reconstruction on a grid $32\times32$ over a $[0,1]\times[0,1]$ domain.

Table \ref{tab:param} shows the values of the parameters for all examples. %All the tests were performed using a $10 \times 10$ pressure sensor network with $1\%$ relative noise. 

\begin{table}[]
	\centering
    \caption{Parameters used in all benchmark tests. \rule[-8pt]{0pt}{12pt} \label{tab:param}}
    \begin{tabular}{ll}
    \hline
    Parameter                        & Value \\ \hline
    Likelihood parameter $a_0$  & 0.001 \\
    Likelihood parameter $b_0$   & 0.001 \\
    Number of particles $N$          & 740   \\
    Number of bridging densities $M$ & 740   \\
    ESS threshold                    & 0.95N \\
    Refinement threshold $\epsilon$  & 0.02  \\
    Mixing factor $\gamma_1$  & 0.85  \\
    Mixing factor $\gamma_s,\quad s>1$  & 0.5  \\
    variance $\alpha_s^{-1}$  & 0.7  \\
    scaling factor $r$  & 0.5  \\
    \hline
    \end{tabular}
    \vspace{12pt}
\end{table}

\FloatBarrier

\begin{figure}[!t] 
    \setlength{\unitlength}{0.01\textwidth} 
    \begin{picture}(100,120)
        % S = 0
        \put(2.5,103.75){\includegraphics[width=0.2\textwidth]{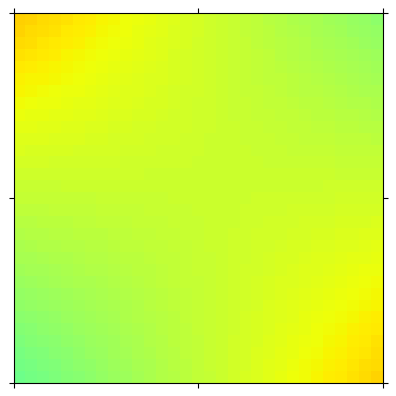}}
        \put(24,103.75){\includegraphics[width=0.2\textwidth]{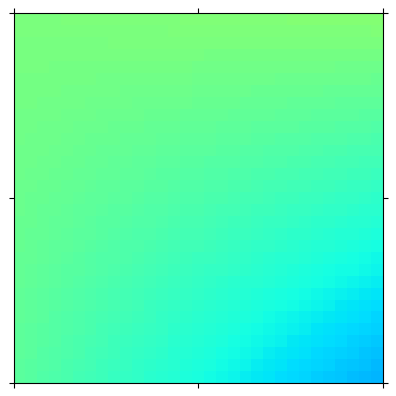}}
        \put(45.5,103.75){\includegraphics[width=0.2\textwidth]{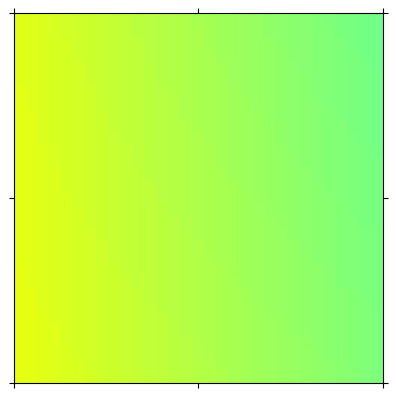}}
        \put(67,103.75){\includegraphics[width=0.2\textwidth]{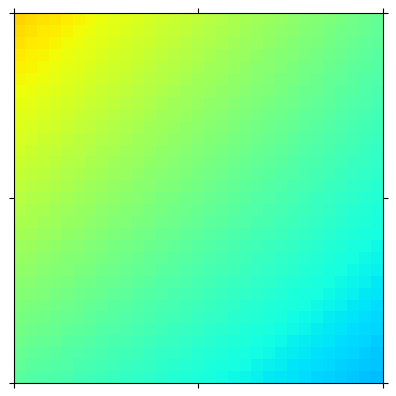}}
        % S = 1
        \put(2.5,83){\includegraphics[width=0.2\textwidth]{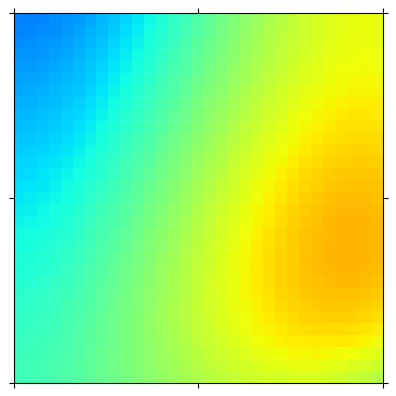}}
        \put(24,83){\includegraphics[width=0.2\textwidth]{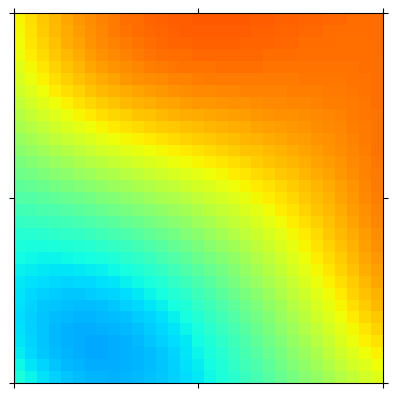}}
        \put(45.5,83){\includegraphics[width=0.2\textwidth]{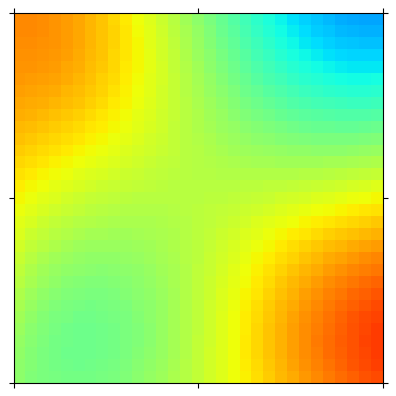}}
        \put(67,83){\includegraphics[width=0.2\textwidth]{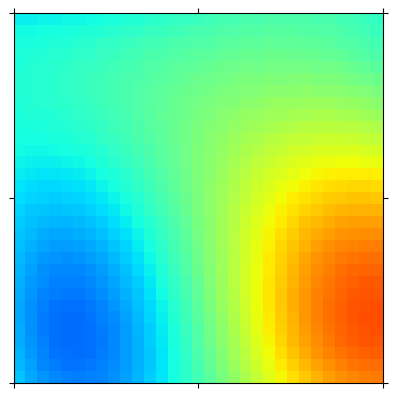}}
        % S = 2
        \put(2.5,62.25){\includegraphics[width=0.2\textwidth]{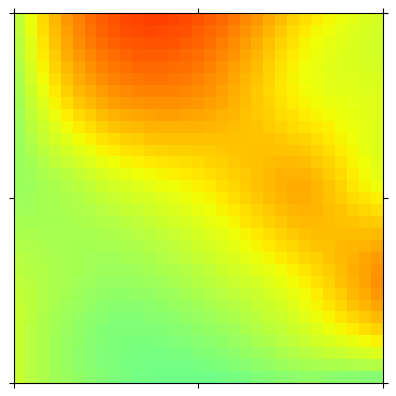}}
        \put(24,62.25){\includegraphics[width=0.2\textwidth]{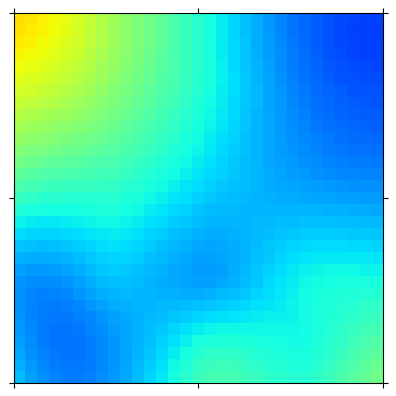}}
        \put(45.5,62.25){\includegraphics[width=0.2\textwidth]{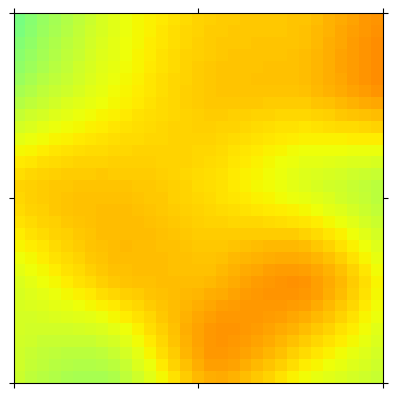}}
        \put(67,62.25){\includegraphics[width=0.2\textwidth]{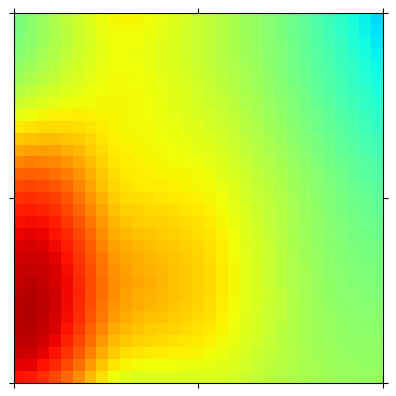}}
        % S = 3
        \put(2.5,41.5){\includegraphics[width=0.2\textwidth]{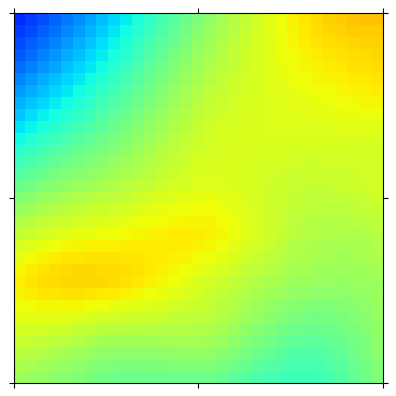}}
        \put(24,41.5){\includegraphics[width=0.2\textwidth]{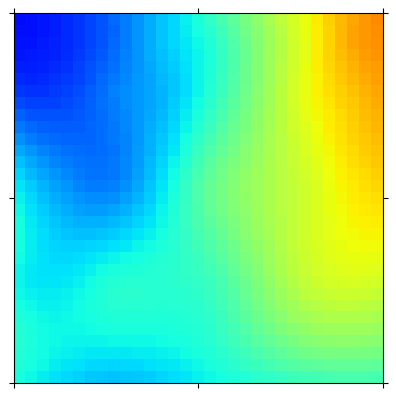}}
        \put(45.5,41.5){\includegraphics[width=0.2\textwidth]{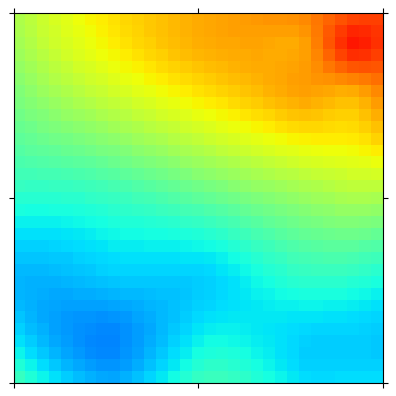}}
        \put(67,41.5){\includegraphics[width=0.2\textwidth]{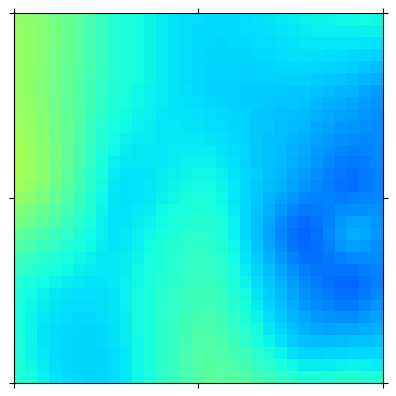}}
        % S = 4
        \put(2.5,20.75){\includegraphics[width=0.2\textwidth]{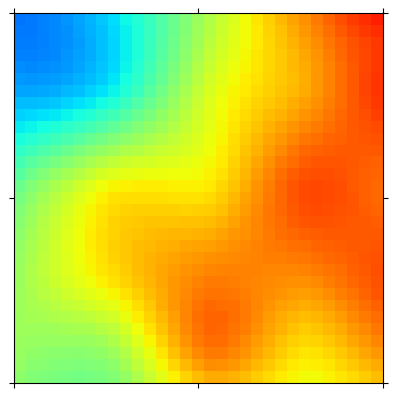}}
        \put(24,20.75){\includegraphics[width=0.2\textwidth]{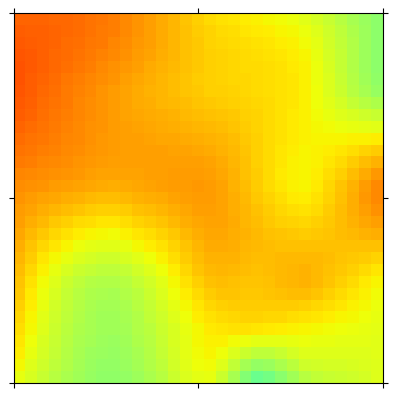}}
        \put(45.5,20.75){\includegraphics[width=0.2\textwidth]{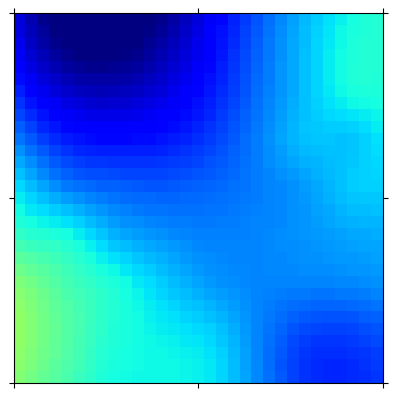}}
        \put(67,20.75){\includegraphics[width=0.2\textwidth]{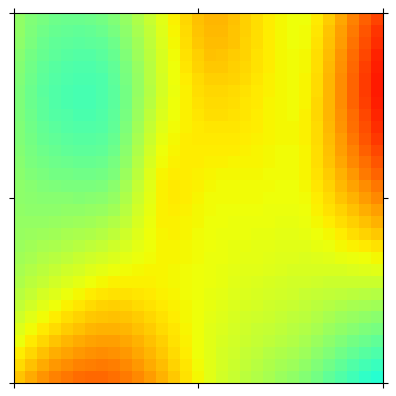}}
        % S = 5
        \put(2.5,0){\includegraphics[width=0.2\textwidth]{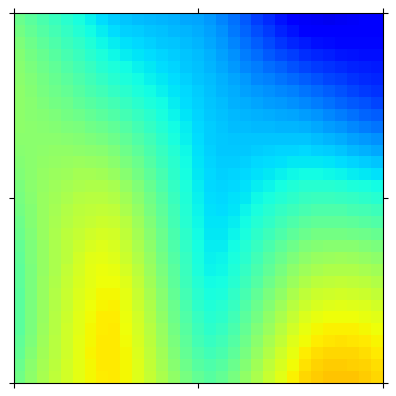}}
        \put(24,0){\includegraphics[width=0.2\textwidth]{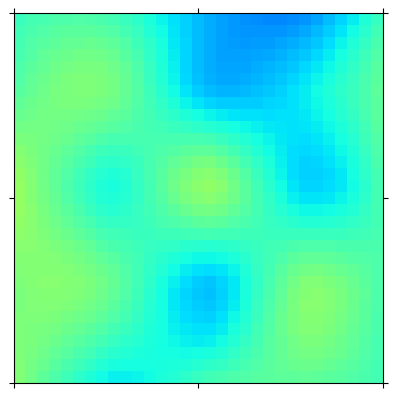}}
        \put(45.5,0){\includegraphics[width=0.2\textwidth]{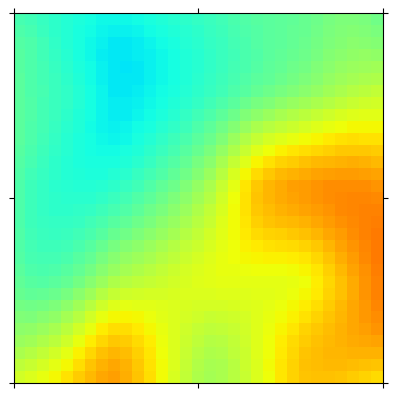}}
        \put(67,0){\includegraphics[width=0.2\textwidth]{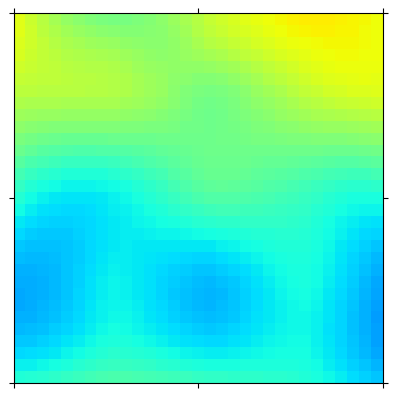}}
        % scale vertical text
        \put(0,7){\rotatebox{90}{\footnotesize $S=5$}}
        \put(0,27.75){\rotatebox{90}{\footnotesize $S=4$}}
        \put(0,48.5){\rotatebox{90}{\footnotesize $S=3$}}
        \put(0,69.25){\rotatebox{90}{\footnotesize $S=2$}}
        \put(0,90){\rotatebox{90}{\footnotesize $S=1$}}
        \put(0,110.75){\rotatebox{90}{\footnotesize $S=0$}}
        % color bar
        \put(90,0.5){\includegraphics[width=0.0453\textwidth]{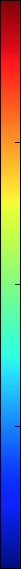}} \put(97,0.5){\large $-2$}
        \put(97,29.5){\large $-1$}
        \put(97,60.5){\large $0$}
        \put(97,91.5){\large $1$}
        \put(97,121.5){\large $2$}
    \end{picture}
    \caption{Log-permeability fields drawn directly from prior distribution for levels $S=0,\dots, 5$. \label{fig:errfig5} }
\end{figure}
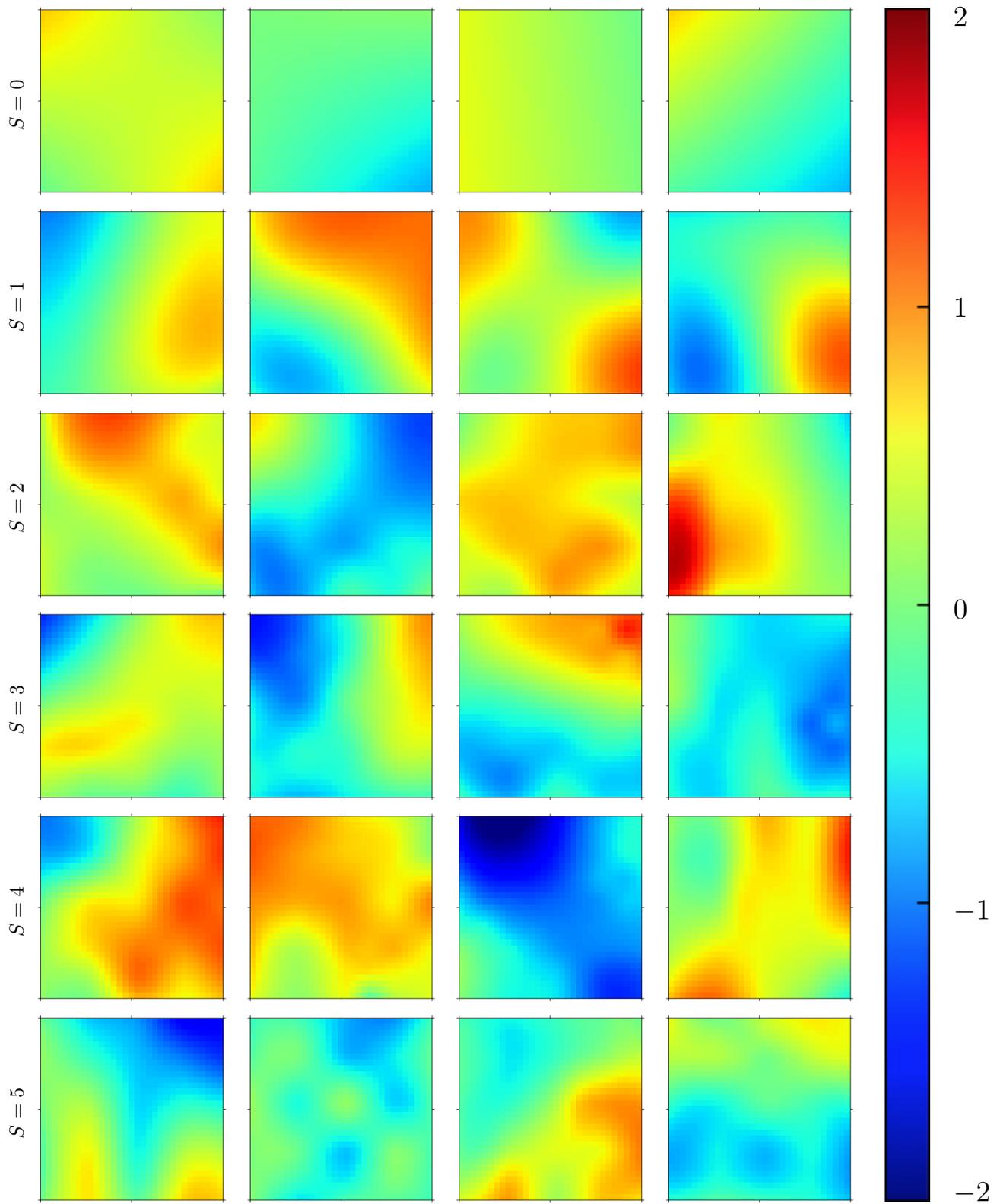
\FloatBarrier

\subsection{Benchmark I: Linear log-permeability field}
In this example, the true log-permeability field has the following form
\begin{equation}
    \text{ln } k(x,y)=2(x+y-1)
\end{equation}
\begin{figure}[h]
	\centering
	\includegraphics[width=5.53cm]{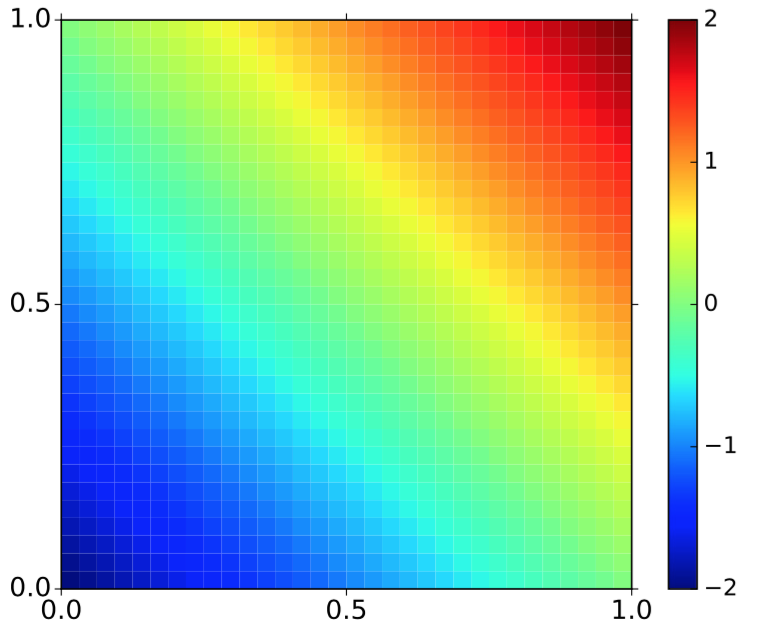}
	\caption{True log-permeability field for benchmark test I.}
\end{figure}
In this test, the algorithm reaches Bayes factor convergence criteria at $S = 4$. Table \ref{tab:baserr} illustrates the number of non-zero bases at different levels of resolution. These compare well with the results of 4, 16, 60, 208, 664 and 1848 reported in \cite{ellam2016bayesian} and show a promotion of sparsity due to the choice of the prior. These results are achieved despite of choosing a lower refinement threshold $\epsilon = 0.02$ in comparison with $\epsilon = 0.05$ used in \cite{ellam2016bayesian}. Note that here the total number of bases for each level of resolution is $2^{2(S+1)}$.
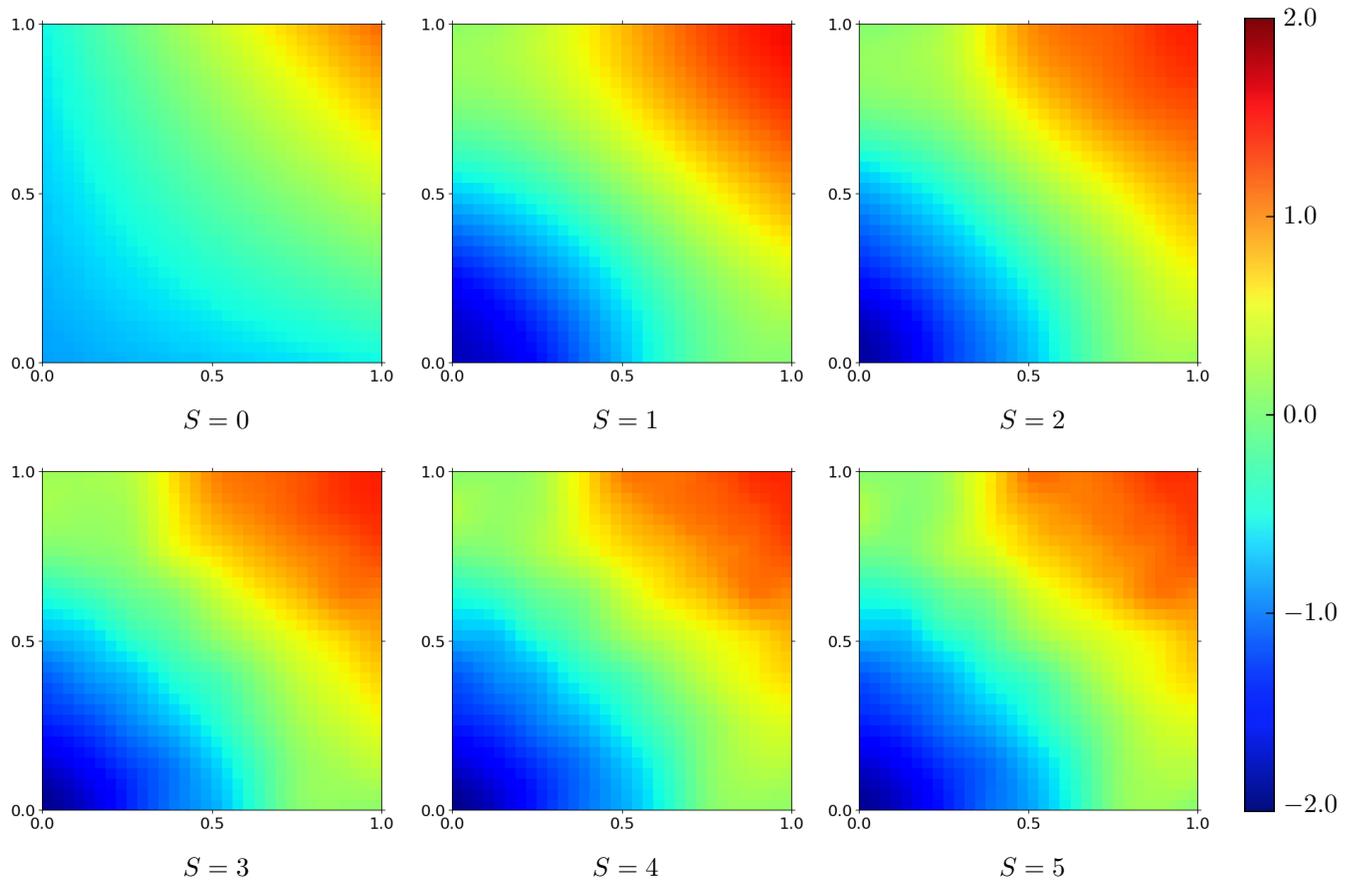
\begin{figure}[h] 
    \setlength{\unitlength}{0.01\textwidth} 
    \begin{picture}(100,67)
        % Top row
        \put(0,36){\includegraphics[width=0.31\textwidth]{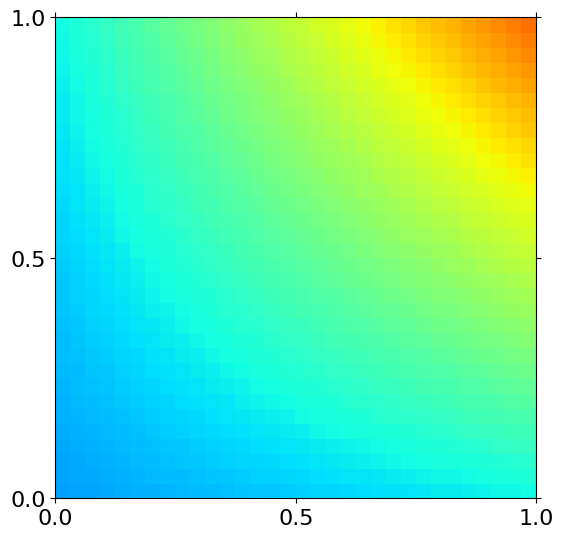}}
        \put(32,36){\includegraphics[width=0.31\textwidth]{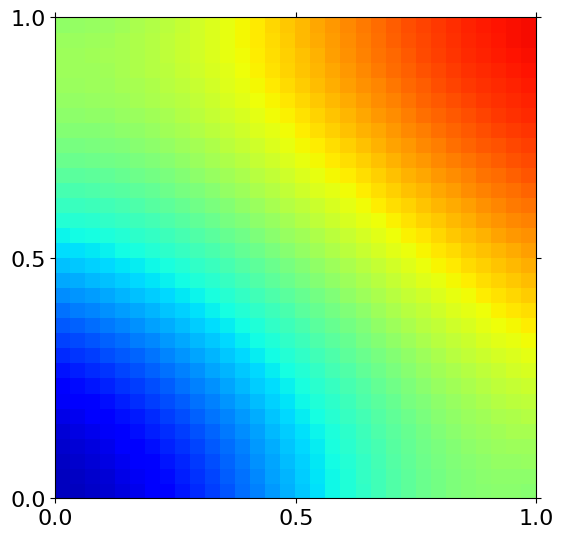}}
        \put(63.8,36){\includegraphics[width=0.31\textwidth]{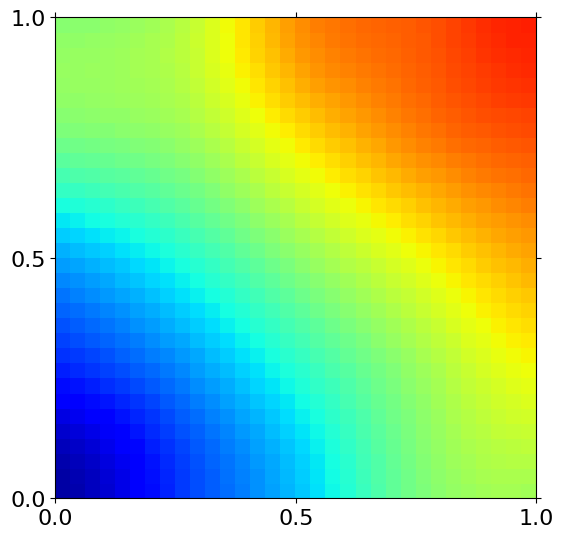}}
        % Bottom row
        \put(0,1){\includegraphics[width=0.31\textwidth]{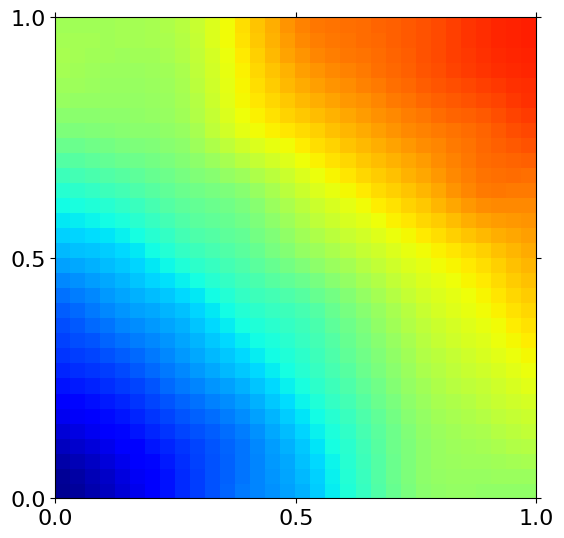}}
        \put(32,1){\includegraphics[width=0.31\textwidth]{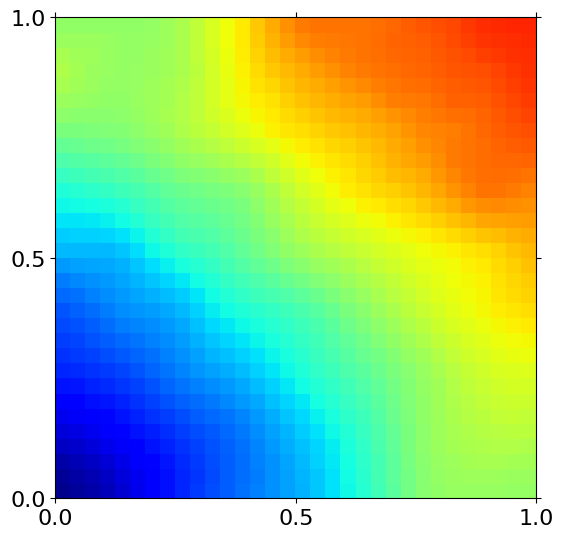}}
        \put(63.8,1){\includegraphics[width=0.31\textwidth]{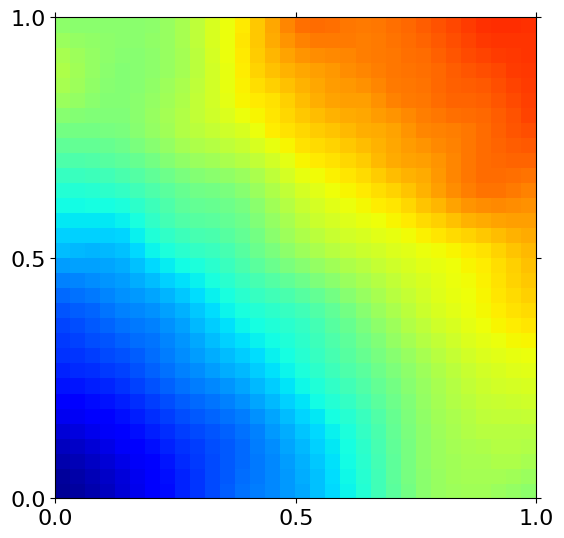}}
        % color bar
        \put(97,3){\includegraphics[width=0.0229\textwidth]{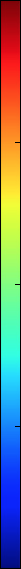}}
        \put(100,64.5){\footnotesize $2.0$}
        \put(100,49){\footnotesize $1.0$}
        \put(100,33.5){\footnotesize $0.0$}
        \put(100,18){\footnotesize $-1.0$}
        \put(100,3){\footnotesize $-2.0$}
        % scale text
        \put(14,33){\footnotesize $S=0$}
        \put(46,33){\footnotesize $S=1$}
        \put(77.8,33){\footnotesize $S=2$}
        \put(14,-2){\footnotesize $S=3$}
        \put(46,-2){\footnotesize $S=4$}
        \put(77.8,-2){\footnotesize $S=5$}
    \end{picture}
    \caption{Posterior means of the log-permeability field of benchmark test I for $S=0,\dots, 5$ with $\%1$ noise and using a $10\times10$ pressure sensor grid. \label{fig:BTIC} }
\end{figure}
\FloatBarrier

\FloatBarrier

\begin{figure}[h] 
    \setlength{\unitlength}{0.01\textwidth} 
    \begin{picture}(100,43)
        % Top row
        % Bottom row
        \put(17.5,-4){\includegraphics[width=0.65\textwidth]{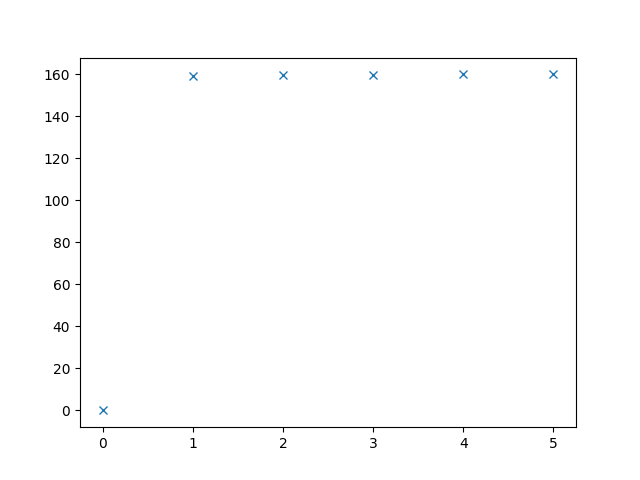}}
        % scale text
        \put(20,16){\rotatebox{90}{\footnotesize $\text{ln } BF_{S,0}$}}
        \put(50,-2.5){\footnotesize $S$}
    \end{picture}
    \caption{Log-Bayes factors for benchmark test I. \label{fig:BFI}}
\end{figure}
\FloatBarrier

\begin{figure}[h]
\centering
% Subfig 1
\subfigure[]{
  \includegraphics[width=5.53cm]{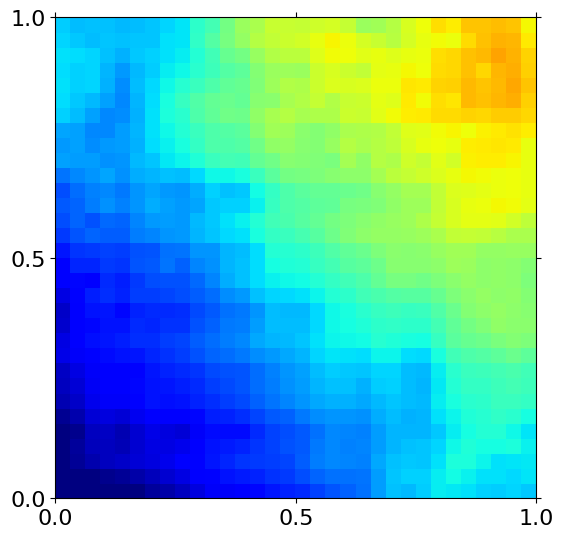}
  \label{subfig:label1}}
% Subfig 2
\subfigure[]{
  \includegraphics[width=5.53cm]{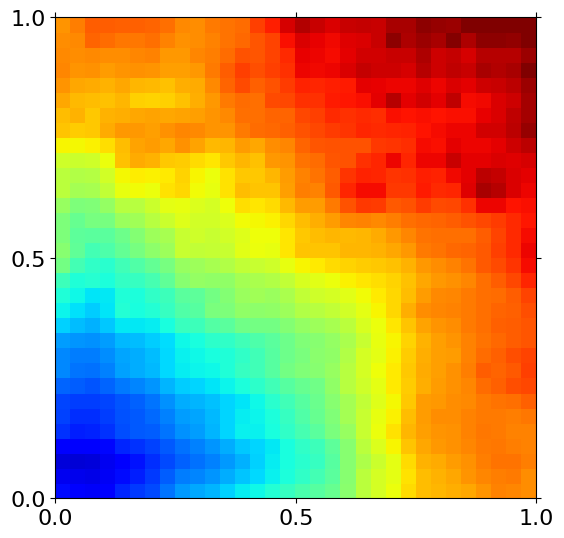}
  \label{subfig:label2}}
\caption{Posterior quantiles of the log-permeability of benchmark test I for $S = 4$ with a $10 \times 10$ sensor network and $1\%$ relative noise. Left shows $5\%$ quantile and right shows $95\%$ quantile.\label{fig:errfig5}}
\end{figure}
\FloatBarrier

\begin{figure}[h]
\centering
% Subfig 1
\subfigure[$S=0$]{
  \includegraphics[width=4.7cm]{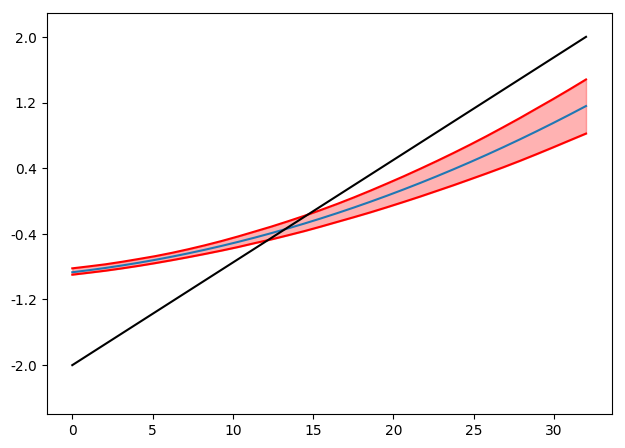}
  \label{subfig:label1}}
% Subfig 2
\subfigure[$S=1$]{
  \includegraphics[width=4.7cm]{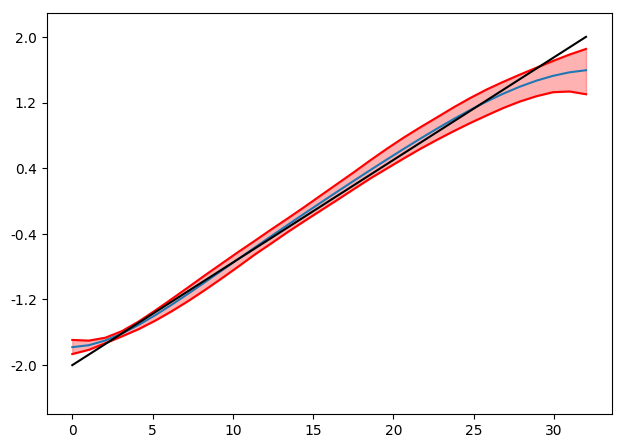}
  \label{subfig:label2}}
% Subfig 3
\subfigure[$S=2$]{
  \includegraphics[width=4.7cm]{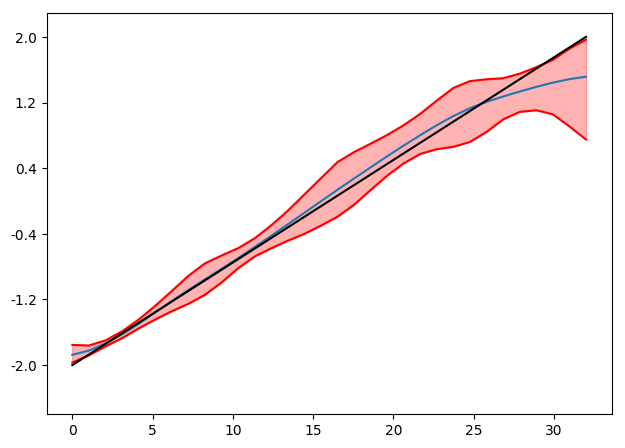}
  \label{subfig:label1}}
  % Subfig 4
\subfigure[$S=3$]{
  \includegraphics[width=4.7cm]{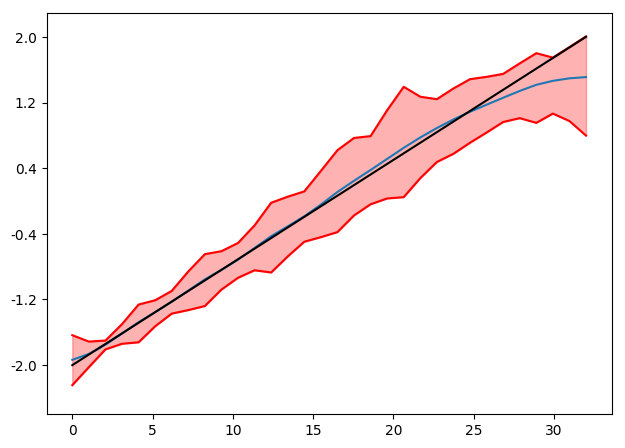}
  \label{subfig:label1}}
% Subfig 5
\subfigure[$S=4$]{
  \includegraphics[width=4.7cm]{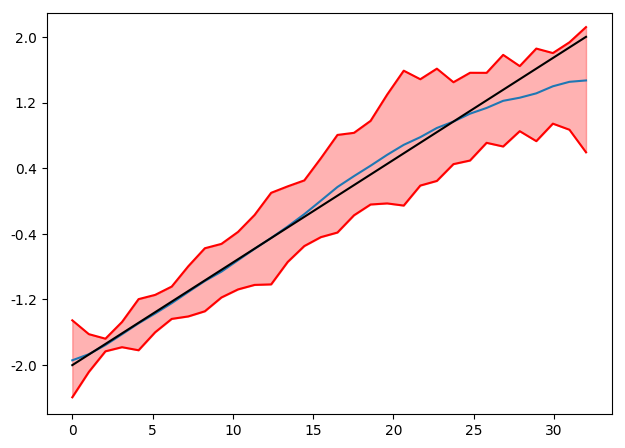}
  \label{subfig:label1}}
% Subfig 6
\subfigure[$S=5$]{
  \includegraphics[width=4.7cm]{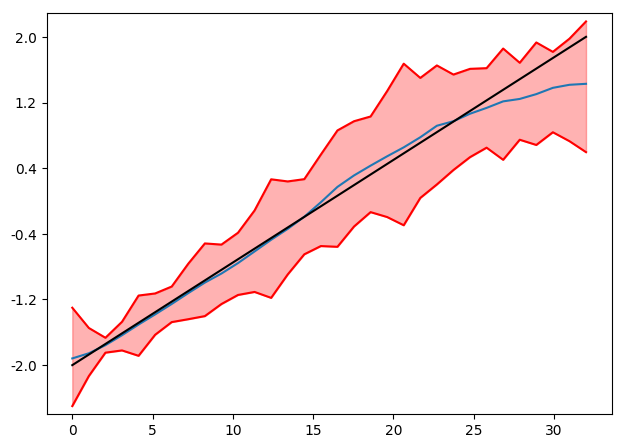}
  \label{subfig:label1}}
\caption{Posterior means of the log-permeability field and $\%90$ credential interval of benchmark test I along the line $x=y$ for $S=0,\dots, 5$ with $\%1$ relative noise and using a $10\times10$ pressure sensor grid. \label{fig:errfig5}}
\end{figure}

\FloatBarrier
\subsection{Benchmark II: Gaussian process realizations}
In this example, the true log-permeability field is the realization of a Gaussian process such that
\begin{equation}
    \text{ln } k(\textbf{x})\sim \mathcal{GP}(m, c(\textbf{x}, \textbf{x}'))
\end{equation}
with zero mean function $m=0$ and an exponential squared covariance function
\begin{equation}
    c(r)=s^2_g\exp{-r^2/2l}
\end{equation}
where $r=\|\textbf{x}- \textbf{x}'\|_2$. Here, the strength is set to $s_g=1$ and length-scale is $l=0.3$. 

In this benchmark test, convergence is not reached until the resolution level $S = 5$. The number of non-zero bases and root mean square error of the converged model is illustrated in Table \ref{tab:baserr}.

\begin{figure}[h]
	\centering
	\includegraphics[width=5.53cm]{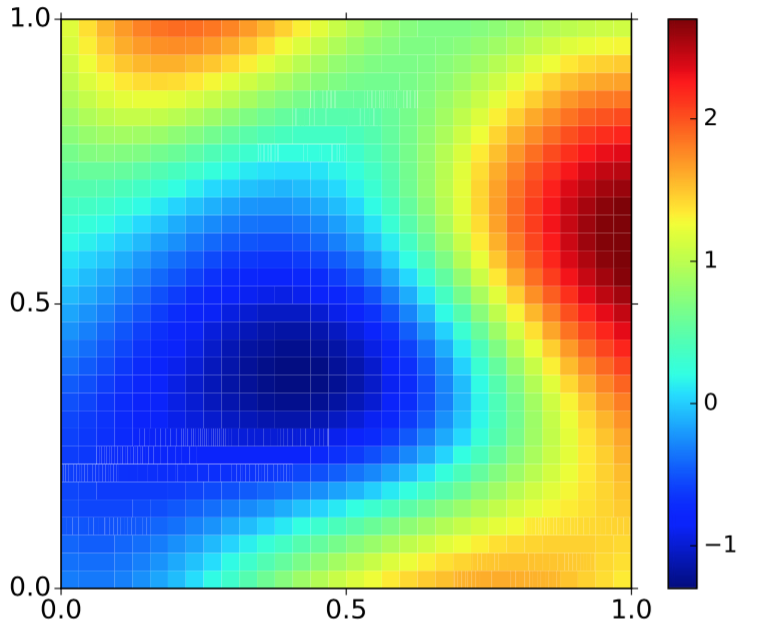}
	\caption{True log-permeability field for benchmark test II.}
\end{figure}

\begin{figure}[h] 
    \setlength{\unitlength}{0.01\textwidth} 
    \begin{picture}(100,67)
        % Top row
        \put(0,36){\includegraphics[width=0.31\textwidth]{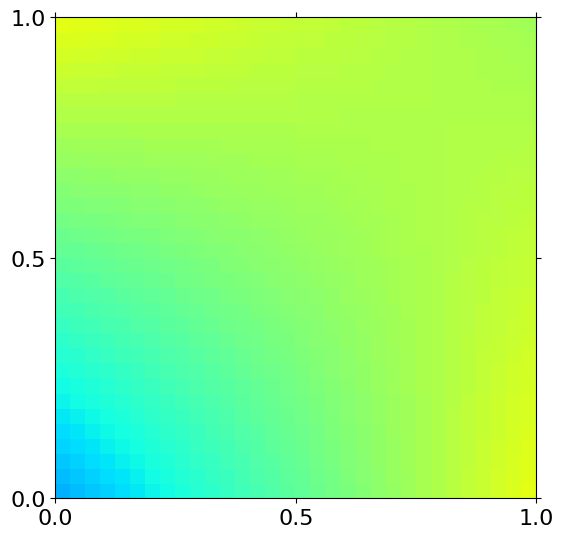}}
        \put(32,36){\includegraphics[width=0.31\textwidth]{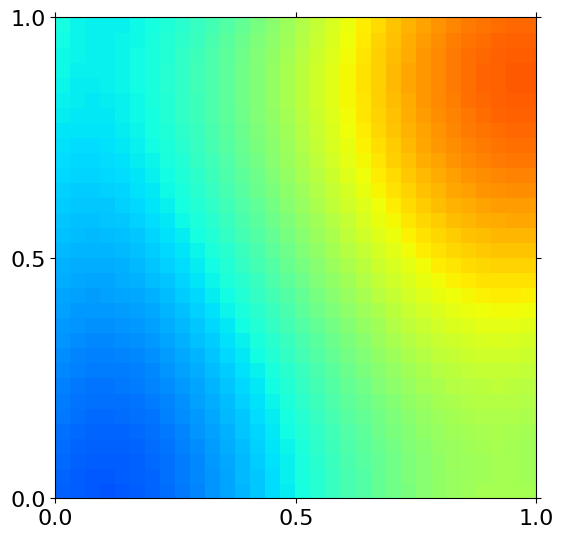}}
        \put(63.8,36){\includegraphics[width=0.31\textwidth]{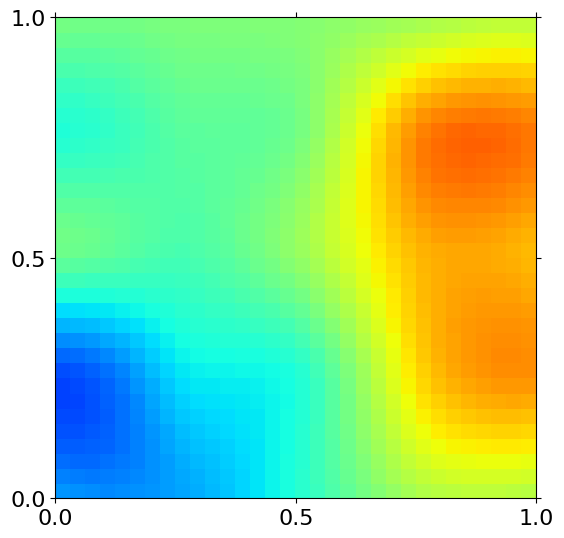}}
        % Bottom row
        \put(0,1){\includegraphics[width=0.31\textwidth]{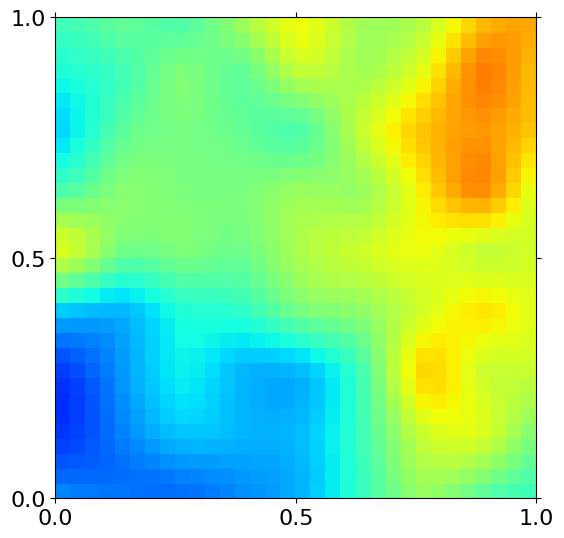}}
        \put(32,1){\includegraphics[width=0.31\textwidth]{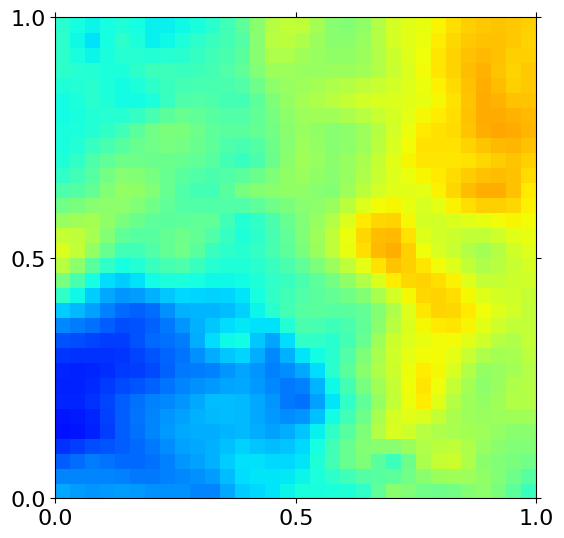}}
        \put(63.8,1){\includegraphics[width=0.31\textwidth]{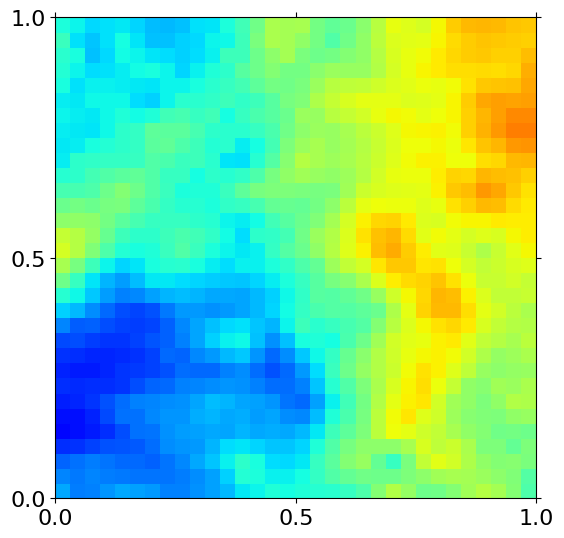}}
        % color bar
        \put(97,3){\includegraphics[width=0.0229\textwidth]{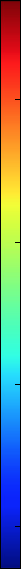}}
        \put(100,53.5){\footnotesize $2.0$}
        \put(100,38){\footnotesize $1.0$}
        \put(100,22.5){\footnotesize $0.0$}
        \put(100,7){\footnotesize $-1.0$}
        % scale text
        \put(14,33){\footnotesize $S=0$}
        \put(46,33){\footnotesize $S=1$}
        \put(77.8,33){\footnotesize $S=2$}
        \put(14,-2){\footnotesize $S=3$}
        \put(46,-2){\footnotesize $S=4$}
        \put(77.8,-2){\footnotesize $S=5$}
    \end{picture}
    \caption{Posterior means of the log-permeability field of benchmark test II for $S=0,\dots, 5$ with $\%1$ relative noise and using a $10\times10$ pressure sensor grid. \label{fig:BTIIC}}
\end{figure}
\FloatBarrier

\begin{figure}[h] 
    \setlength{\unitlength}{0.01\textwidth} 
    \begin{picture}(100,43)
        % Top row
        % Bottom row
        \put(17.5,-4){\includegraphics[width=0.65\textwidth]{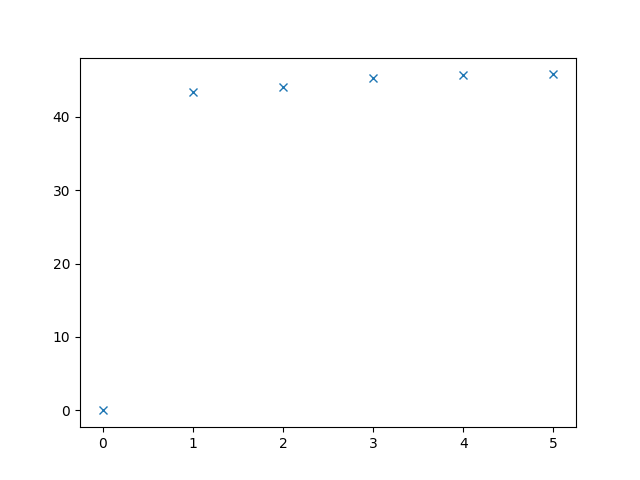}}
        % scale text
        \put(20,16){\rotatebox{90}{\footnotesize $\text{ln } BF_{S,0}$}}
        \put(50,-2.5){\footnotesize $S$}
    \end{picture}
    \caption{Log-Bayes factors for benchmark test II with $\%1$ relative noise using a $10\times10$ pressure sensor network. \label{fig:BFIII}}
\end{figure}
\FloatBarrier

\begin{figure}[h]
\centering
% Subfig 1
\subfigure[]{
  \includegraphics[width=5.53cm]{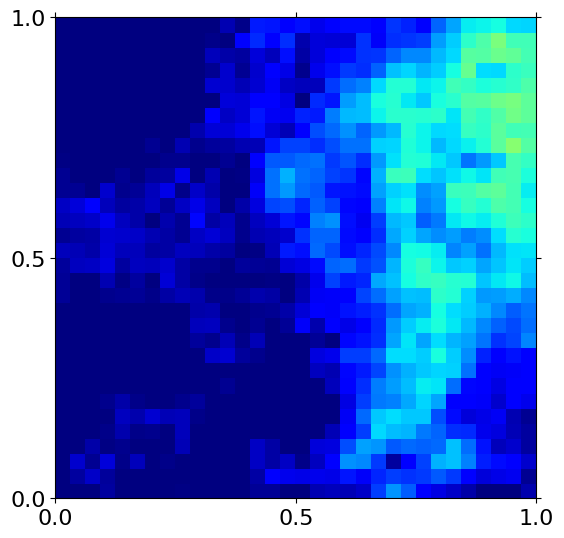}
  \label{subfig:label1}}
% Subfig 2
\subfigure[]{
  \includegraphics[width=5.53cm]{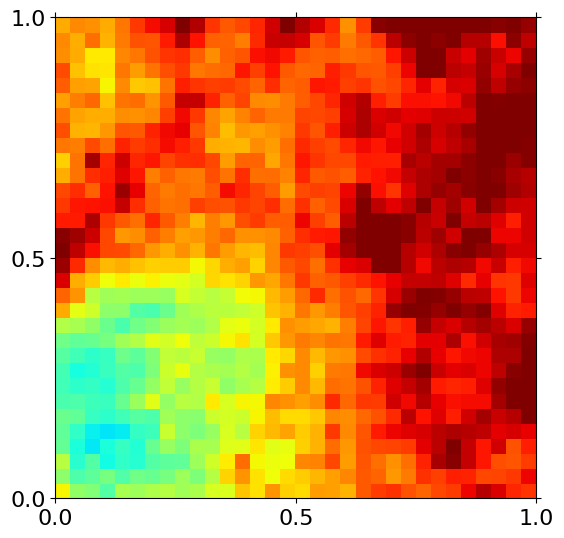}
  \label{subfig:label2}}
\caption{Posterior quantiles of the log-permeability of benchmark test II for $S = 5$ using a $10 \times 10$ sensor network with $1\%$ relative noise. Left shows $5\%$ quantile and right shows $95\%$ quantile. \label{fig:errfig5}}
\end{figure}
\FloatBarrier

\begin{figure}[h]
\centering
% Subfig 1
\subfigure[$S=0$]{
  \includegraphics[width=4.7cm]{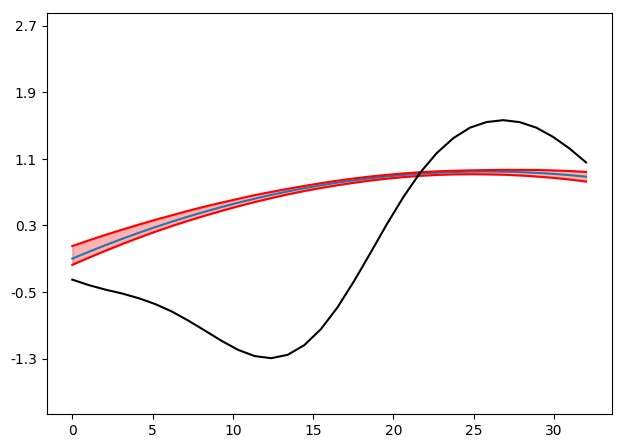}
  \label{subfig:label1}}
% Subfig 2
\subfigure[$S=1$]{
  \includegraphics[width=4.7cm]{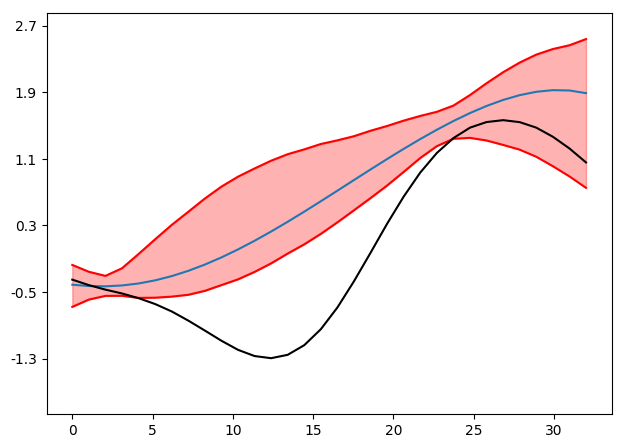}
  \label{subfig:label2}}
% Subfig 3
\subfigure[$S=2$]{
  \includegraphics[width=4.7cm]{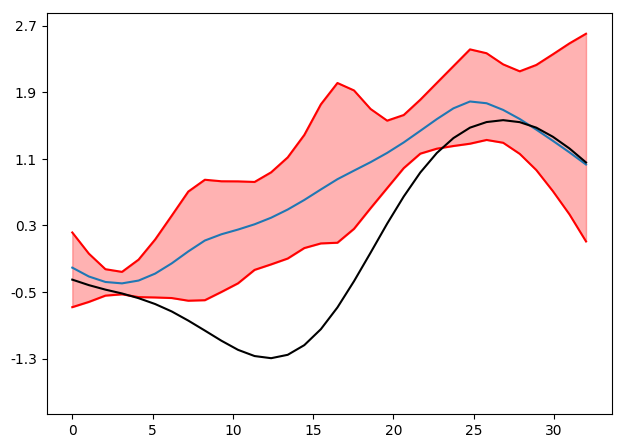}
  \label{subfig:label1}}
  % Subfig 4
\subfigure[$S=3$]{
  \includegraphics[width=4.7cm]{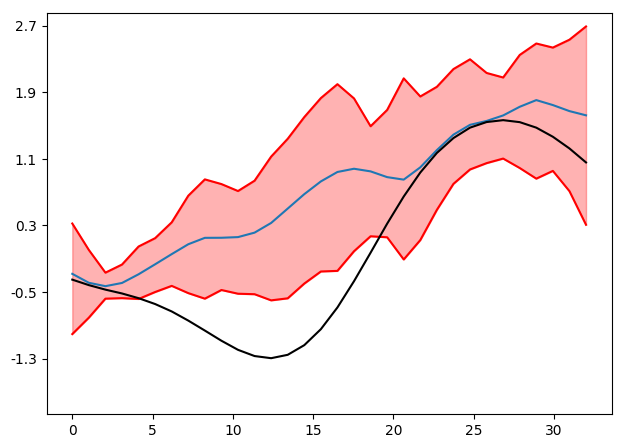}
  \label{subfig:label1}}
% Subfig 5
\subfigure[$S=4$]{
  \includegraphics[width=4.7cm]{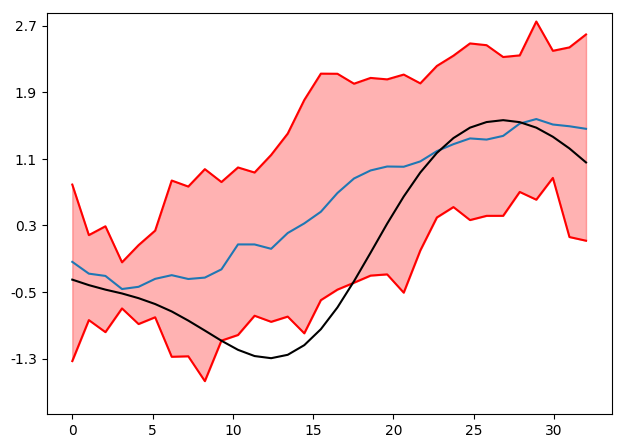}
  \label{subfig:label1}}
% Subfig 6
\subfigure[$S=5$]{
  \includegraphics[width=4.7cm]{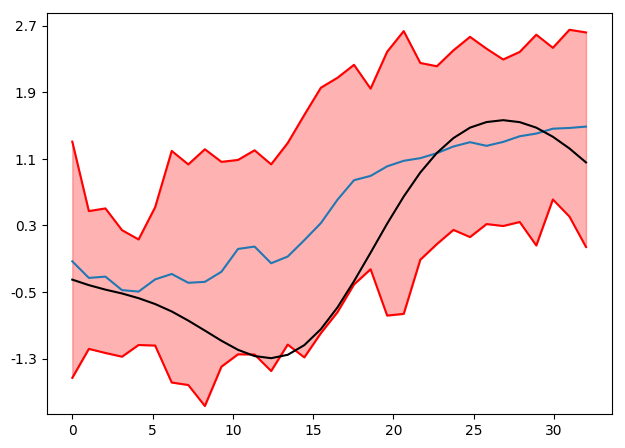}
  \label{subfig:label1}}
\caption{Posterior means of the log-permeability field and $\%90$ credential interval of benchmark test II along the line $x=y$ for $S=0,\dots, 5$ with $\%1$ relative noise and using a $10\times10$ pressure sensor grid.\label{fig:errfig5}}
\end{figure}

\subsection{Benchmark III: Gaussian process realizations}
In this example, the true log-permeability field is the realization of a Gaussian process such that
\begin{equation}
    \text{ln } k(\textbf{x})\sim \mathcal{GP}(m, c(\textbf{x}, \textbf{x}'))
\end{equation}
with zero mean function $m=0$ and an exponential covariance function
\begin{equation}
    c(r)=s^2_g\exp{-r/l}
\end{equation}
where $r=\|\textbf{x}- \textbf{x}'\|_2$. The strength is $s_g=1$ and length-scale is $l=0.3$.

Table \ref{tab:baserr} shows a more sparse solution in comparison with the number of non-zero bases 4, 16, 60, 208, 664, and 1368, reported in \cite{ellam2016bayesian}, for $S = 0,\dots,5$. This test case is converged at $S=3$.

\begin{figure}[h]
	\centering
	\includegraphics[width=5.53cm]{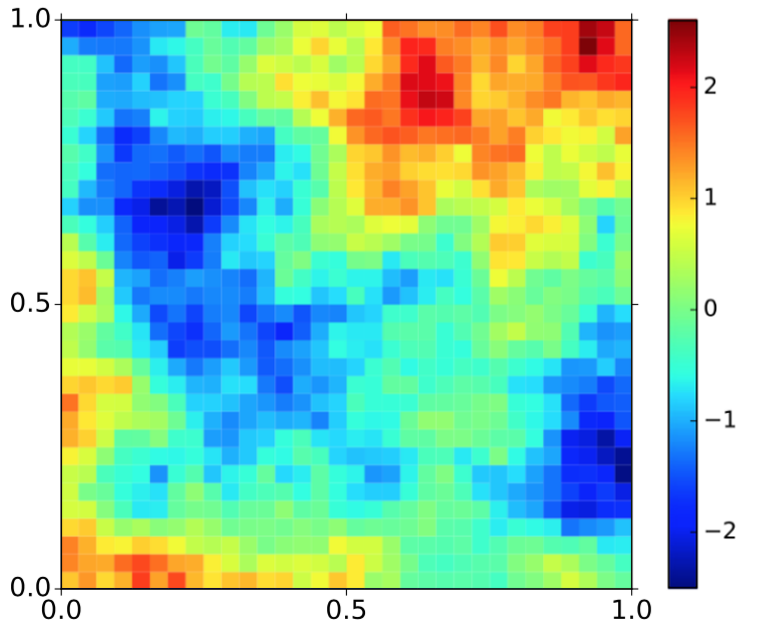}
	\caption{True log-permeability field for benchmark test III.}
\end{figure}

\begin{figure}[h] 
    \setlength{\unitlength}{0.01\textwidth} 
    \begin{picture}(100,67)
        % Top row
        \put(0,36){\includegraphics[width=0.31\textwidth]{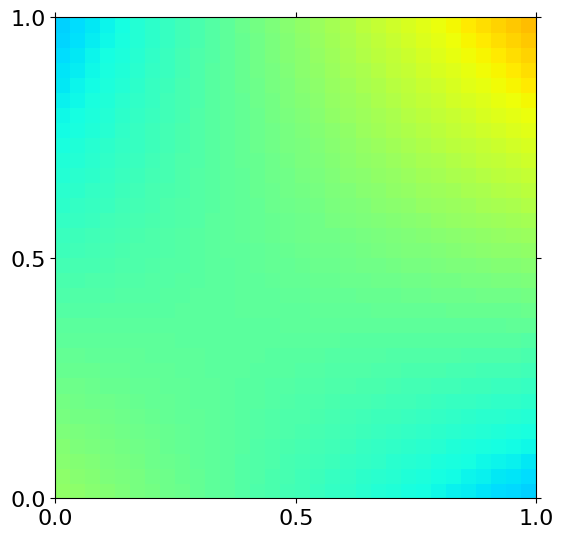}}
        \put(32,36){\includegraphics[width=0.31\textwidth]{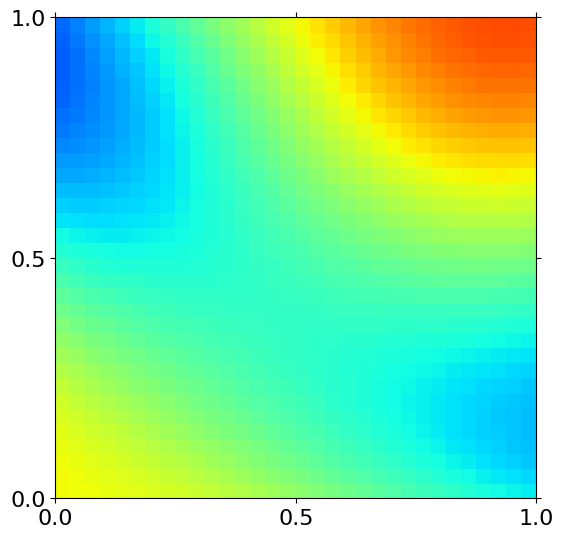}}
        \put(63.8,36){\includegraphics[width=0.31\textwidth]{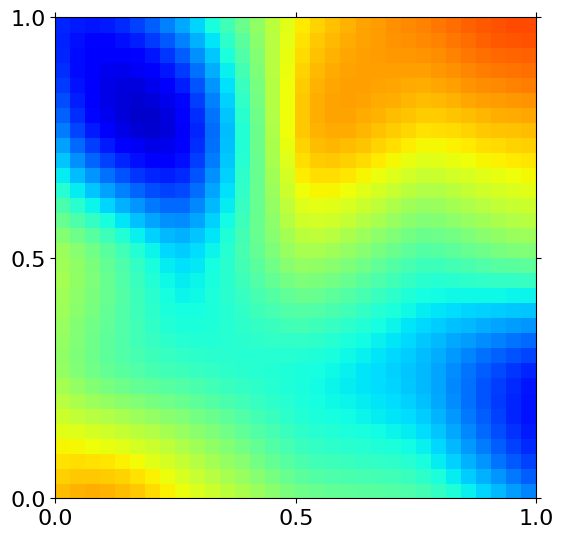}}
        % Bottom row
        \put(0,1){\includegraphics[width=0.31\textwidth]{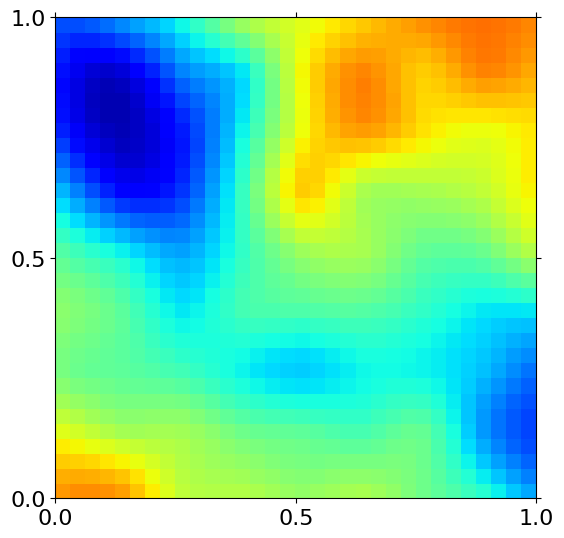}}
        \put(32,1){\includegraphics[width=0.31\textwidth]{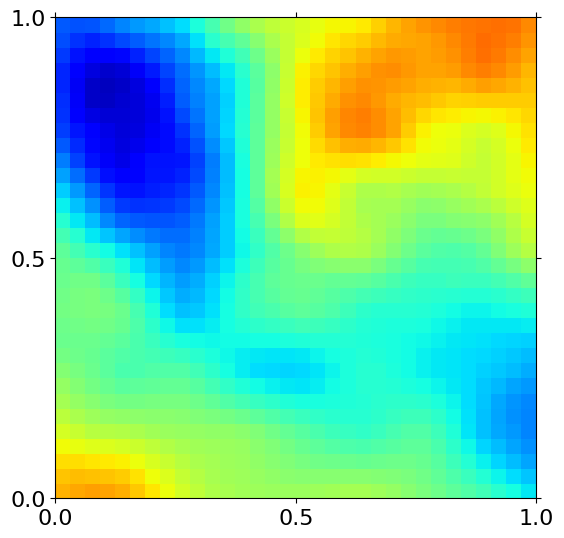}}
        \put(63.8,1){\includegraphics[width=0.31\textwidth]{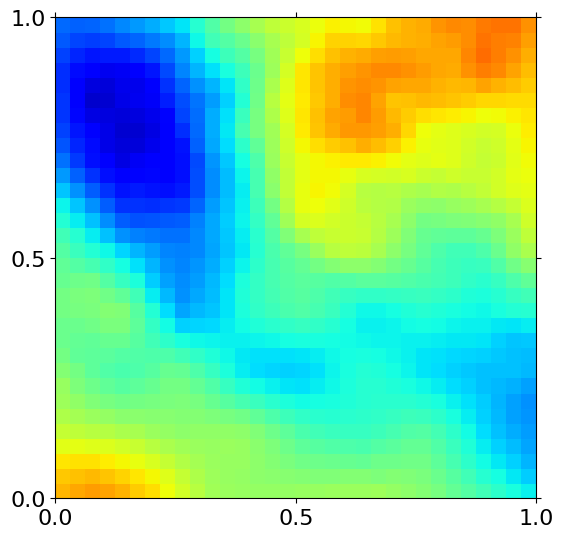}}
        % color bar
        \put(97,3){\includegraphics[width=0.0229\textwidth]{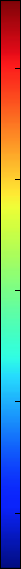}}
        \put(100,57){\footnotesize $2.0$}
        \put(100,45){\footnotesize $1.0$}
        \put(100,33){\footnotesize $0.0$}
        \put(100,21){\footnotesize $-1.0$}
        \put(100,9){\footnotesize $-2.0$}
        % scale text
        \put(14,33){\footnotesize $S=0$}
        \put(46,33){\footnotesize $S=1$}
        \put(77.8,33){\footnotesize $S=2$}
        \put(14,-2){\footnotesize $S=3$}
        \put(46,-2){\footnotesize $S=4$}
        \put(77.8,-2){\footnotesize $S=5$}
    \end{picture}
    \caption{Posterior means of the log-permeability field of benchmark test III for $S=0,\dots, 5$ with $\%1$ relative noise and using a $10\times10$ pressure sensor grid. \label{fig:BTIIIC}}
\end{figure}
\FloatBarrier

\begin{figure}[h] 
    \setlength{\unitlength}{0.01\textwidth} 
    \begin{picture}(100,43)
        % Top row
        % Bottom row
        \put(17.5,-4){\includegraphics[width=0.65\textwidth]{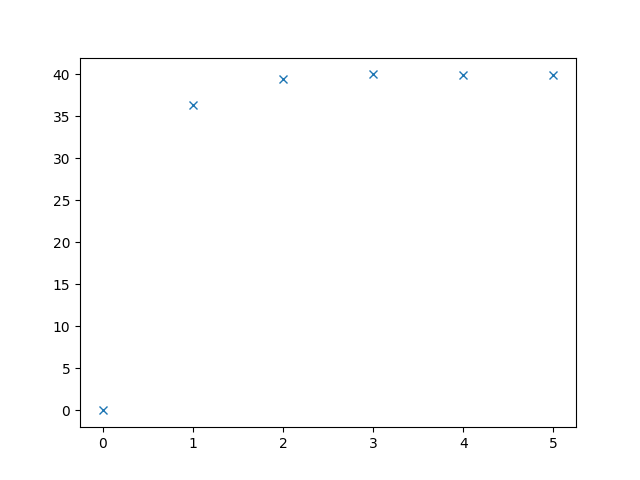}}
        % scale text
        \put(20,16){\rotatebox{90}{\footnotesize $\text{ln } BF_{S,0}$}}
        \put(50,-2.5){\footnotesize $S$}
    \end{picture}
    \caption{Log-Bayes factors for benchmark test III with $\%1$ relative noise using a $10\times10$ pressure sensor network. \label{fig:BFIII}}
\end{figure}
\FloatBarrier

\begin{figure}[h]
\centering
% Subfig 1
\subfigure[]{
  \includegraphics[width=5.53cm]{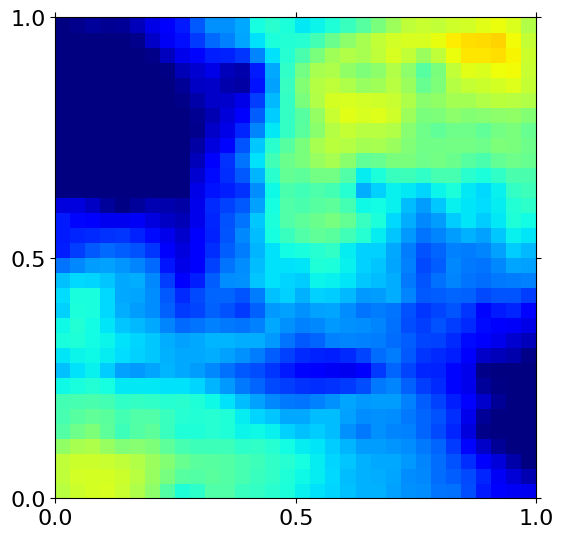}
  \label{subfig:label1}}
% Subfig 2
\subfigure[]{
  \includegraphics[width=5.53cm]{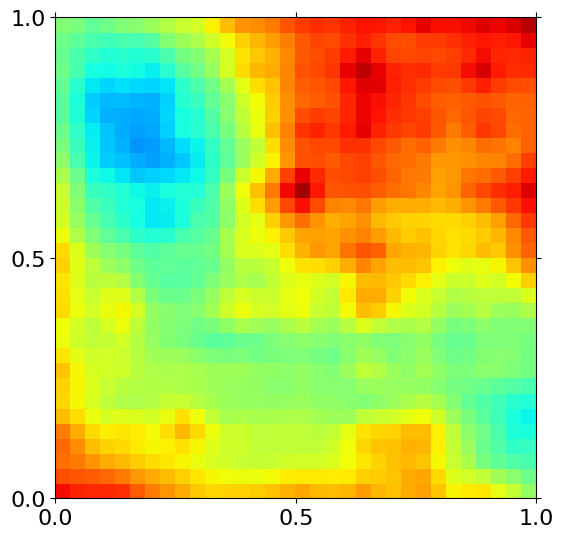}
  \label{subfig:label2}}
\caption{Posterior quantiles of the log-permeability of benchmark test III for $S = 3$ using a $10 \times 10$ pressure sensor grid with $1\%$ relative noise. Left shows $5\%$ quantile and right shows $95\%$ quantile.\label{fig:errfig5}}
\end{figure}
\FloatBarrier

\begin{figure}[h]
\centering
% Subfig 1
\subfigure[$S=0$]{
  \includegraphics[width=4.7cm]{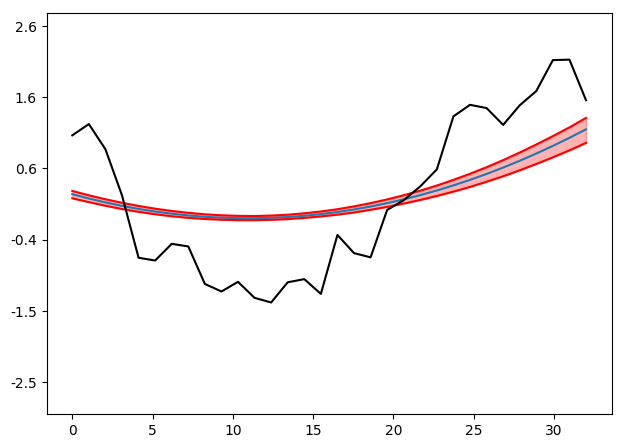}
  \label{subfig:label1}}
% Subfig 2
\subfigure[$S=1$]{
  \includegraphics[width=4.7cm]{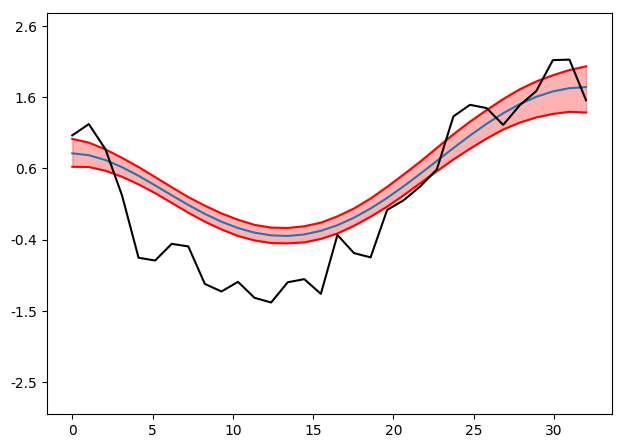}
  \label{subfig:label2}}
% Subfig 3
\subfigure[$S=2$]{
  \includegraphics[width=4.7cm]{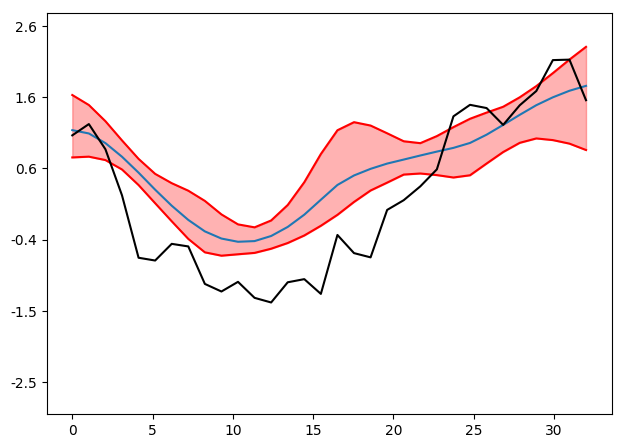}
  \label{subfig:label1}}
  % Subfig 4
\subfigure[$S=3$]{
  \includegraphics[width=4.7cm]{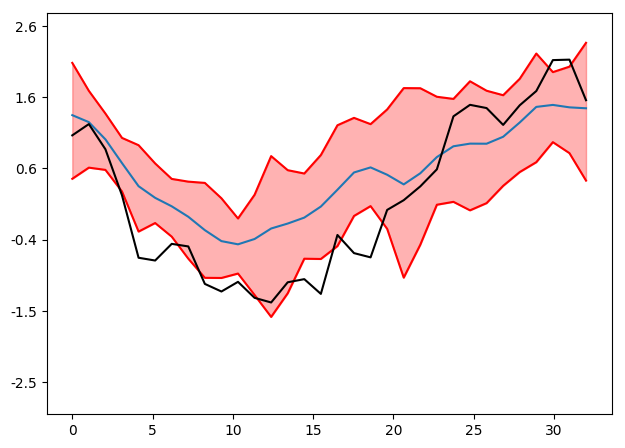}
  \label{subfig:label1}}
% Subfig 5
\subfigure[$S=4$]{
  \includegraphics[width=4.7cm]{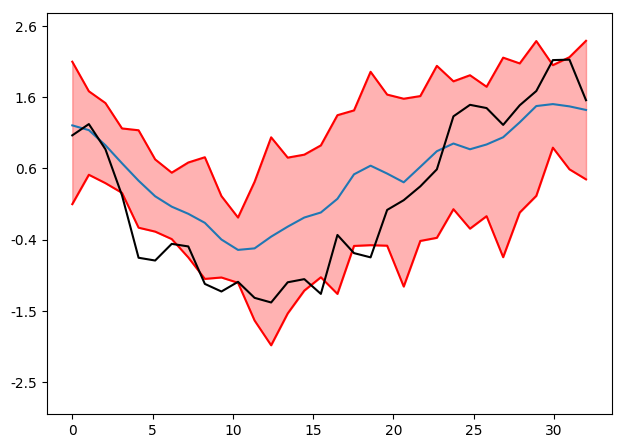}
  \label{subfig:label1}}
% Subfig 6
\subfigure[$S=5$]{
  \includegraphics[width=4.7cm]{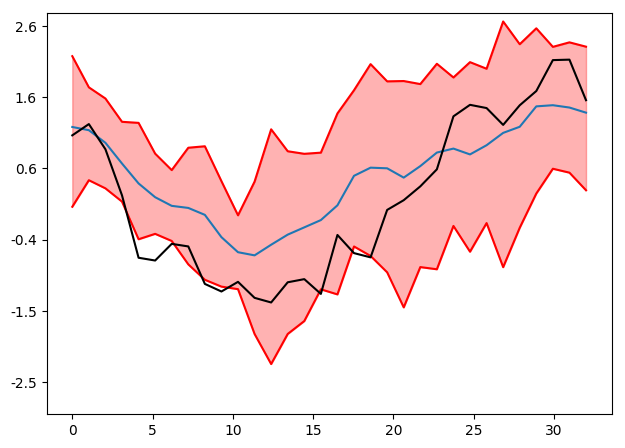}
  \label{subfig:label1}}
\caption{Posterior means of the log-permeability field and $\%90$ credential interval of benchmark test III along the line $x=y$ for $S=0,\dots, 5$ with $\%1$ relative noise and using a $10\times10$ pressure sensor grid. \label{fig:errfig5}}
\end{figure}

\begin{figure}[h] 
    \setlength{\unitlength}{0.01\textwidth} 
    \begin{picture}(100,67)
        % Top row
        \put(0,36){\includegraphics[width=0.31\textwidth]{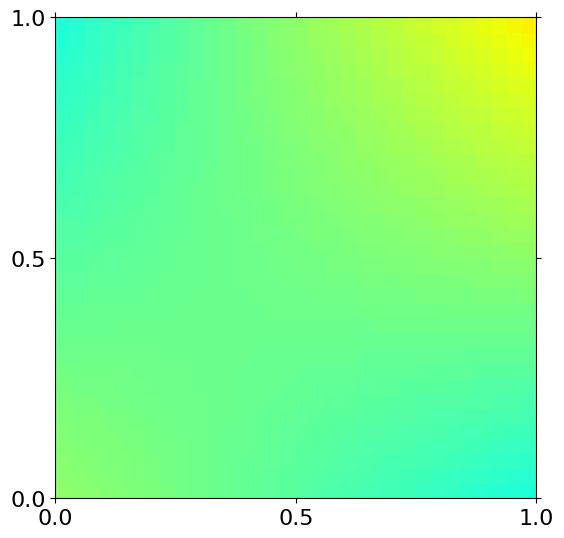}}
        \put(32,36){\includegraphics[width=0.31\textwidth]{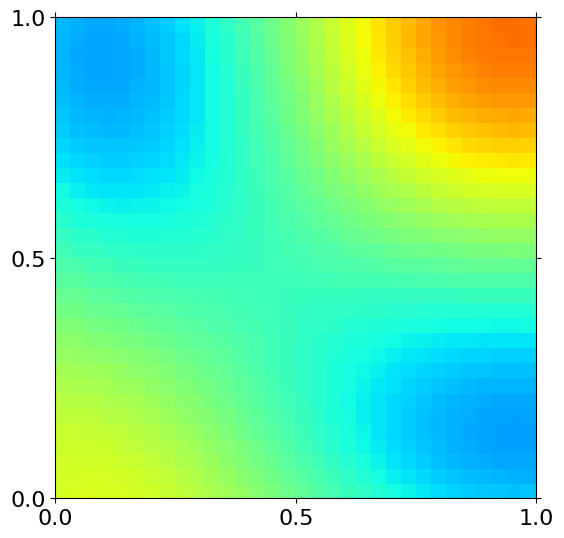}}
        \put(63.8,36){\includegraphics[width=0.31\textwidth]{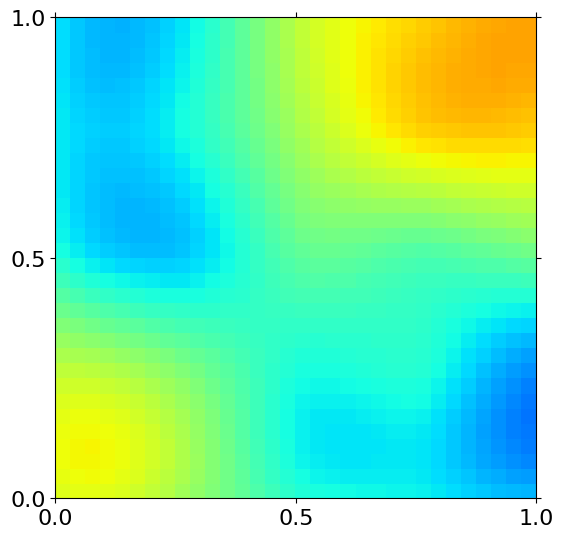}}
        % Bottom row
        \put(0,1){\includegraphics[width=0.31\textwidth]{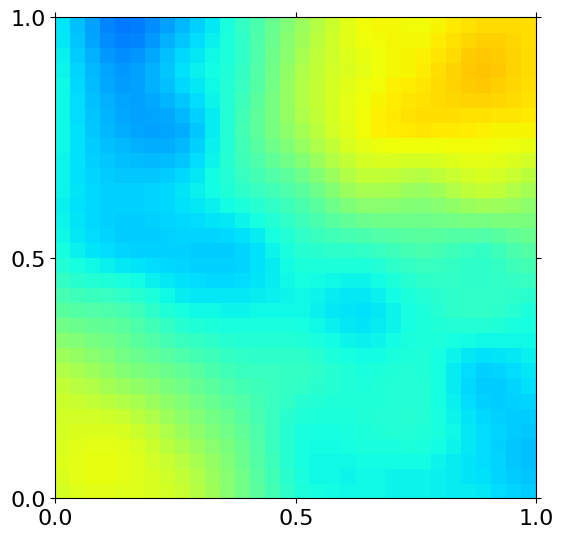}}
        \put(32,1){\includegraphics[width=0.31\textwidth]{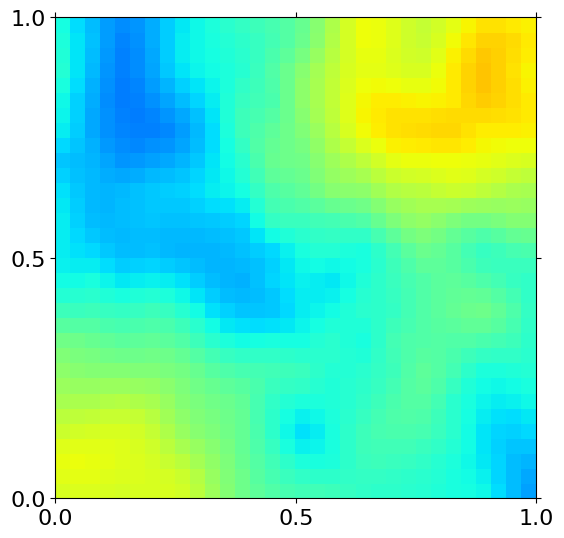}}
        \put(63.8,1){\includegraphics[width=0.31\textwidth]{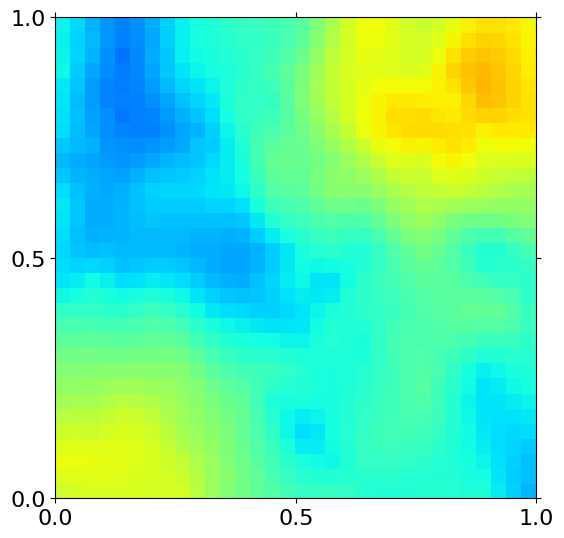}}
        % color bar
        \put(97,3){\includegraphics[width=0.0229\textwidth]{figures/caseC/10x10/caseC_colorbar.png}}
        \put(100,57){\footnotesize $2.0$}
        \put(100,45){\footnotesize $1.0$}
        \put(100,33){\footnotesize $0.0$}
        \put(100,21){\footnotesize $-1.0$}
        \put(100,9){\footnotesize $-2.0$}
        % scale text
        \put(14,33){\footnotesize $S=0$}
        \put(46,33){\footnotesize $S=1$}
        \put(77.8,33){\footnotesize $S=2$}
        \put(14,-2){\footnotesize $S=3$}
        \put(46,-2){\footnotesize $S=4$}
        \put(77.8,-2){\footnotesize $S=5$}
    \end{picture}
    \caption{Posterior means of the log-permeability field of benchmark test III for $S=0,\dots, 5$ using a $5\times5$ pressure sensor grid and with $\%1$ relative noise. \label{fig:BTIII5xC}}
\end{figure}
\FloatBarrier

% \begin{figure}[h] 
%     \setlength{\unitlength}{0.01\textwidth} 
%     \begin{picture}(100,45)
%         % Top row
%         % Bottom row
%         \put(17.5,-4){\includegraphics[width=0.65\textwidth]{figures/caseC/5x5/BF.png}}
%         % scale text
%         \put(21,16){\rotatebox{90}{\footnotesize $\text{ln } BF_{S,0}$}}
%         \put(50,-2.5){\footnotesize $S$}
%     \end{picture}
%     \caption{Log-Bayes factors for benchmark test III with $\%1$ relative noise using a $5\times5$ pressure sensor grid. \label{fig:BFIII5x5}}
% \end{figure}
% \FloatBarrier

\begin{figure}[h]
\centering
% Subfig 1
\subfigure[$S=0$]{
  \includegraphics[width=4.7cm]{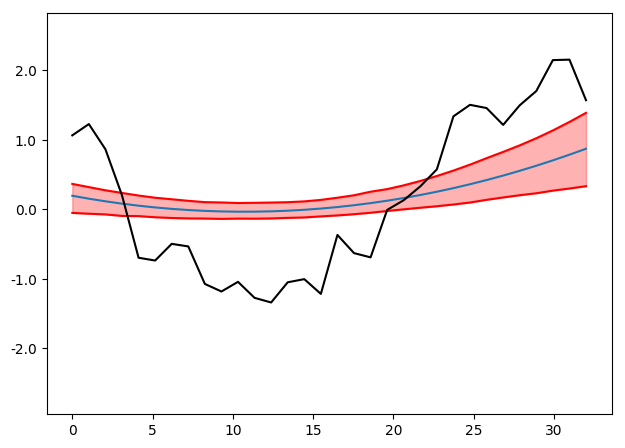}
  \label{subfig:label1}}
% Subfig 2
\subfigure[$S=1$]{
  \includegraphics[width=4.7cm]{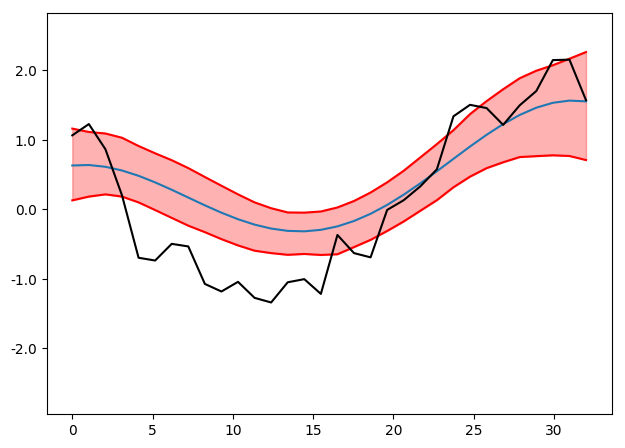}
  \label{subfig:label2}}
% Subfig 3
\subfigure[$S=2$]{
  \includegraphics[width=4.7cm]{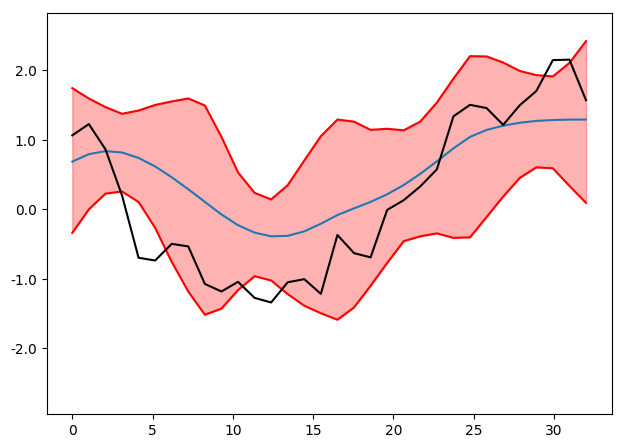}
  \label{subfig:label1}}
  % Subfig 4
\subfigure[$S=3$]{
  \includegraphics[width=4.7cm]{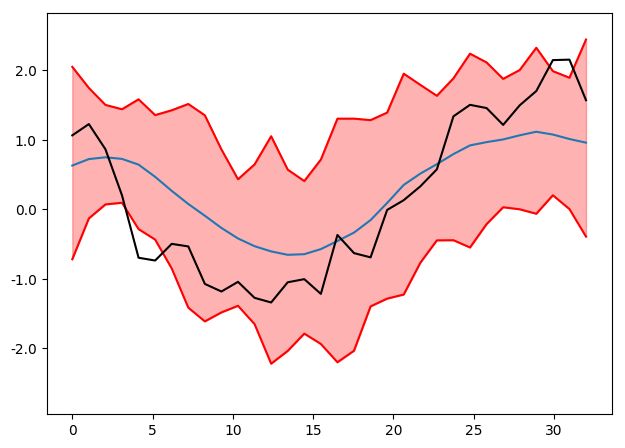}
  \label{subfig:label1}}
% Subfig 5
\subfigure[$S=4$]{
  \includegraphics[width=4.7cm]{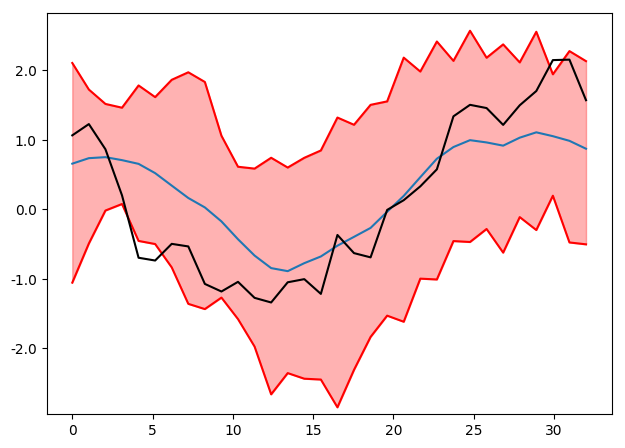}
  \label{subfig:label1}}
% Subfig 6
\subfigure[$S=5$]{
  \includegraphics[width=4.7cm]{figures/caseC/5x5/bar_level4.png}
  \label{subfig:label1}}
\caption{Posterior means of the log-permeability field of benchmark test III along the line $x=y$ for $S=0,\dots, 5$ with $\%1$ relative noise and using a $5\times5$ pressure sensor grid. \label{fig:errfig5}}
\end{figure}

\begin{figure}[h] 
    \setlength{\unitlength}{0.01\textwidth} 
    \begin{picture}(100,67)
        % Top row
        \put(0,36){\includegraphics[width=0.31\textwidth]{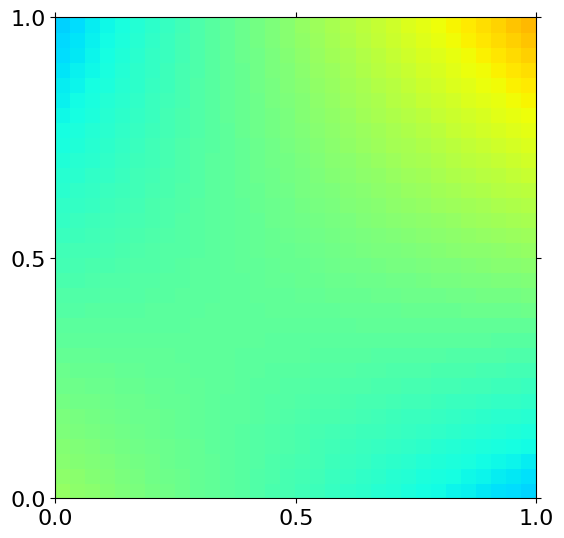}}
        \put(32,36){\includegraphics[width=0.31\textwidth]{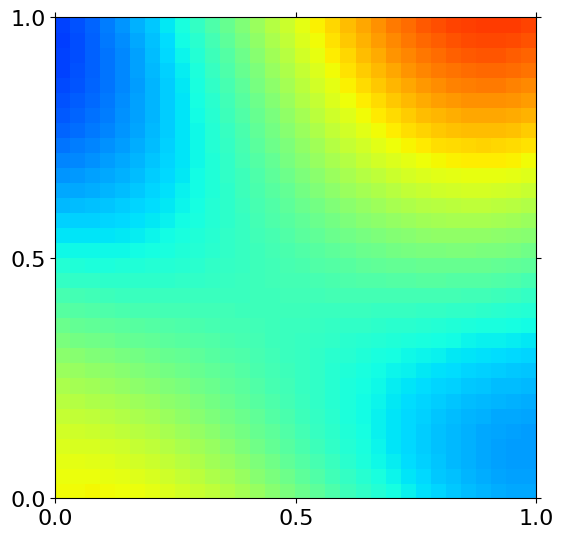}}
        \put(63.8,36){\includegraphics[width=0.31\textwidth]{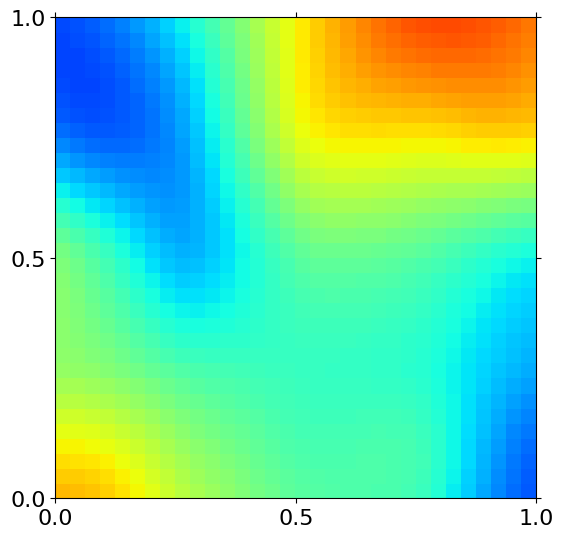}}
        % Bottom row
        \put(0,1){\includegraphics[width=0.31\textwidth]{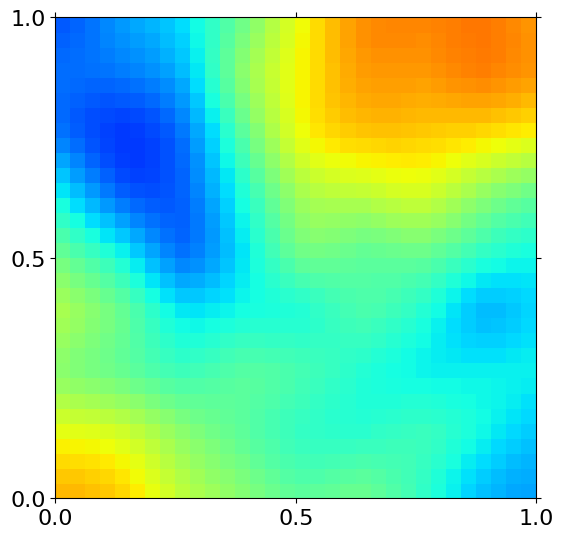}}
        \put(32,1){\includegraphics[width=0.31\textwidth]{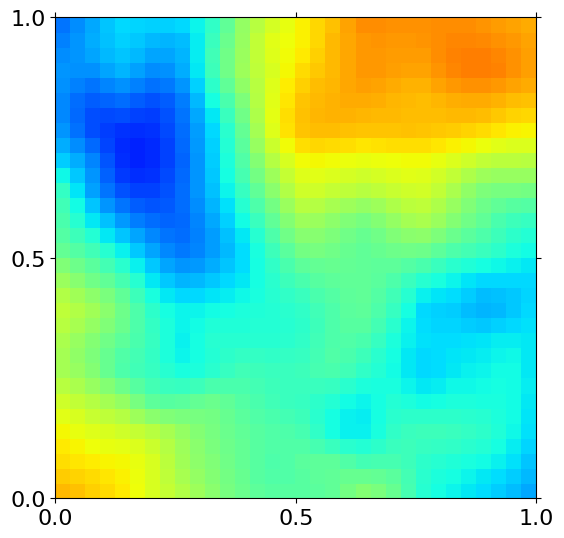}}
        \put(63.8,1){\includegraphics[width=0.31\textwidth]{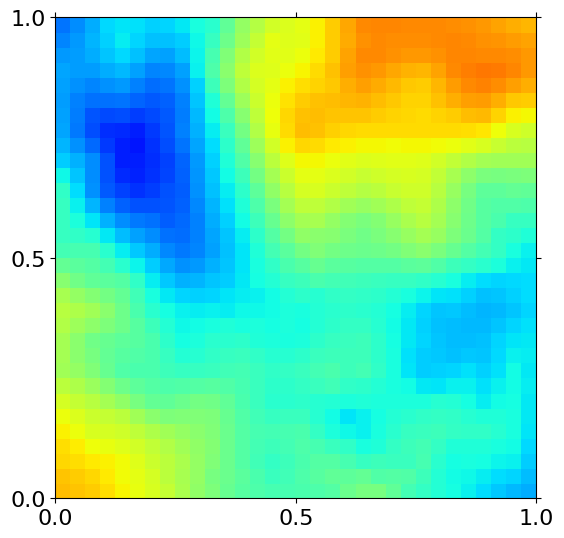}}
        % color bar
        \put(97,3){\includegraphics[width=0.0229\textwidth]{figures/caseC/10x10/caseC_colorbar.png}}
        \put(100,57){\footnotesize $2.0$}
        \put(100,45){\footnotesize $1.0$}
        \put(100,33){\footnotesize $0.0$}
        \put(100,21){\footnotesize $-1.0$}
        \put(100,9){\footnotesize $-2.0$}
        % scale text
        \put(14,33){\footnotesize $S=0$}
        \put(46,33){\footnotesize $S=1$}
        \put(77.8,33){\footnotesize $S=2$}
        \put(14,-2){\footnotesize $S=3$}
        \put(46,-2){\footnotesize $S=4$}
        \put(77.8,-2){\footnotesize $S=5$}
    \end{picture}
    \caption{Posterior means of the log-permeability field of benchmark test III for $S=0,\dots, 5$ with $\%5$ relative noise and using a $10\times10$ pressure sensor grid. \label{fig:BTIII5pC}}
\end{figure}
\FloatBarrier

\begin{figure}[h]
\centering
% Subfig 1
\subfigure[$S=0$]{
  \includegraphics[width=4.7cm]{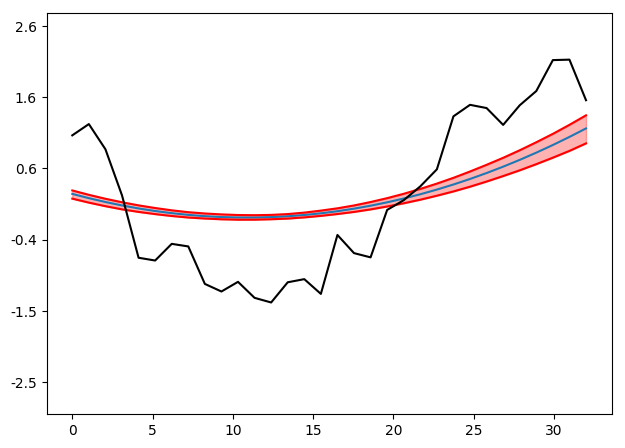}
  \label{subfig:label1}}
% Subfig 2
\subfigure[$S=1$]{
  \includegraphics[width=4.7cm]{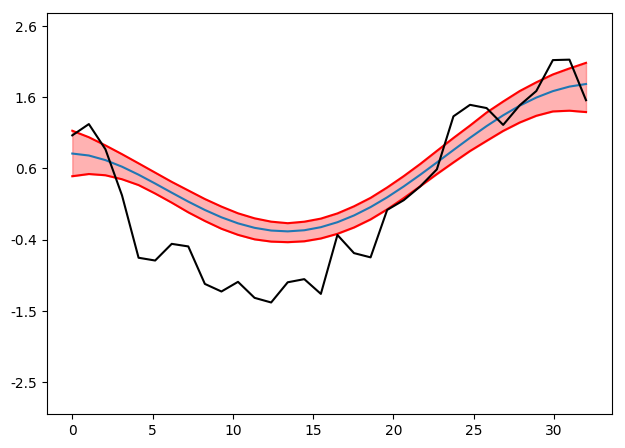}
  \label{subfig:label2}}
% Subfig 3
\subfigure[$S=2$]{
  \includegraphics[width=4.7cm]{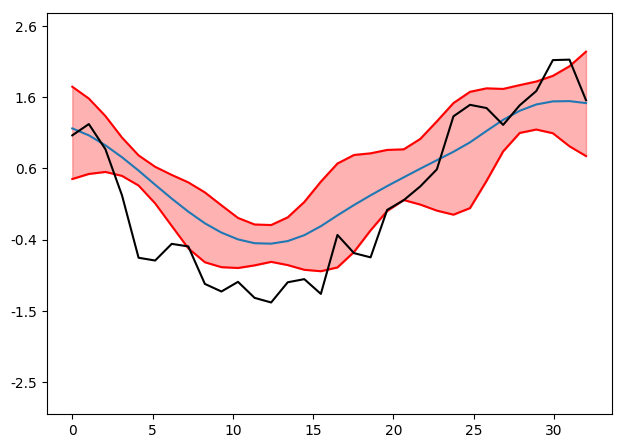}
  \label{subfig:label1}}
  % Subfig 4
\subfigure[$S=3$]{
  \includegraphics[width=4.7cm]{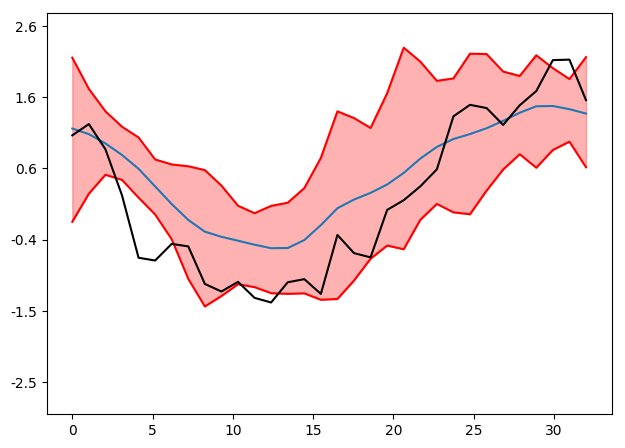}
  \label{subfig:label1}}
% Subfig 5
\subfigure[$S=4$]{
  \includegraphics[width=4.7cm]{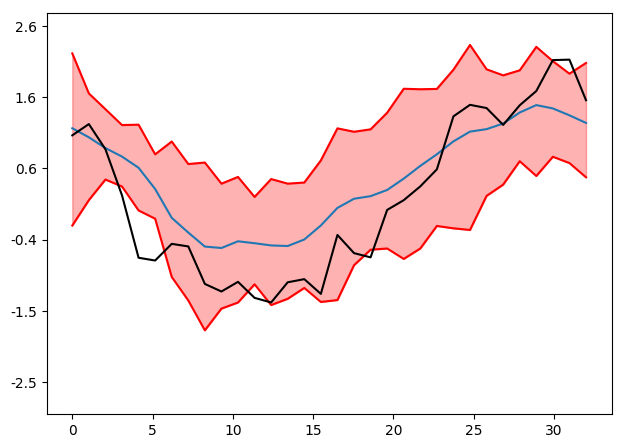}
  \label{subfig:label1}}
% Subfig 6
\subfigure[$S=5$]{
  \includegraphics[width=4.7cm]{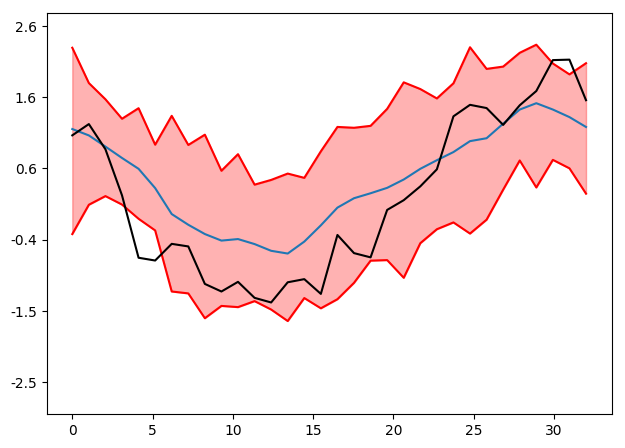}
  \label{subfig:label1}}
\caption{Posterior means of the log-permeability field of benchmark test III along the line $x=y$ for $S=0,\dots, 5$ with $\%5$ relative noise and using a $10\times10$ pressure sensor grid.\label{fig:errfig5}}
\end{figure}

\begin{figure}[h] 
    \setlength{\unitlength}{0.01\textwidth} 
    \begin{picture}(105,55)
        % Top row
        \put(1,29){\includegraphics[width=0.31\textwidth]{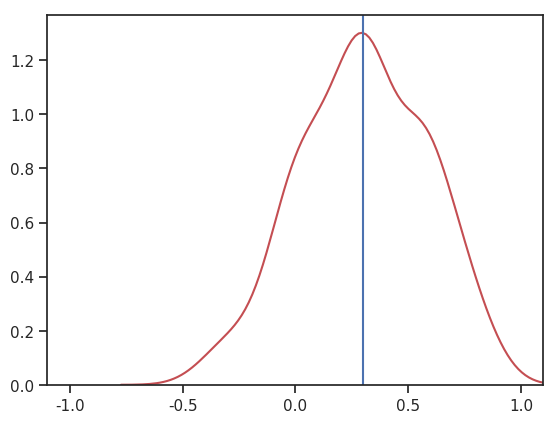}}
        \put(36,29){\includegraphics[width=0.31\textwidth]{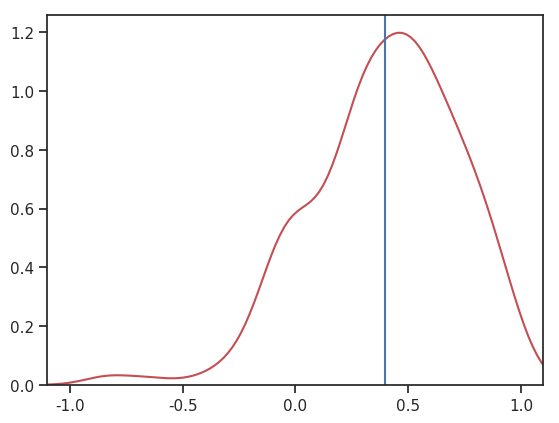}}
        \put(70.8,29){\includegraphics[width=0.31\textwidth]{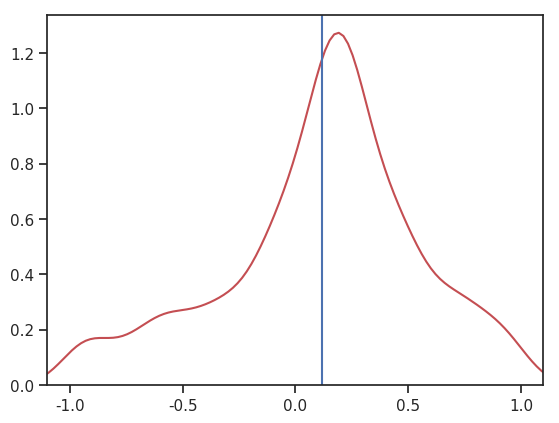}}
        % Bottom row
        \put(1,0){\includegraphics[width=0.31\textwidth]{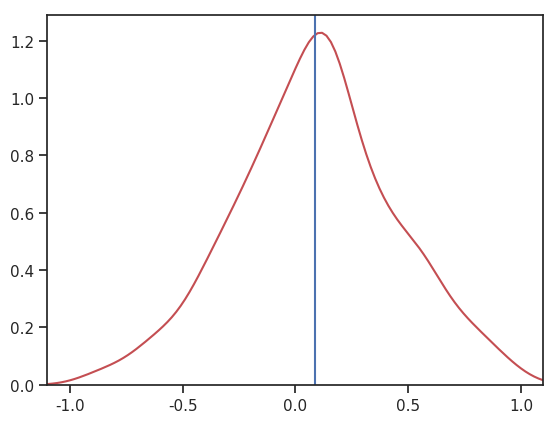}}
        \put(36,0){\includegraphics[width=0.31\textwidth]{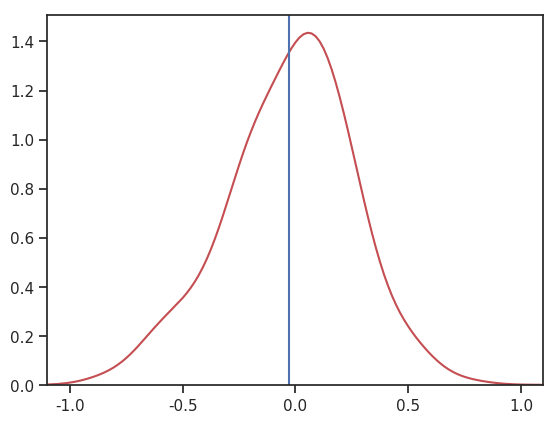}}
        \put(70.8,0){\includegraphics[width=0.31\textwidth]{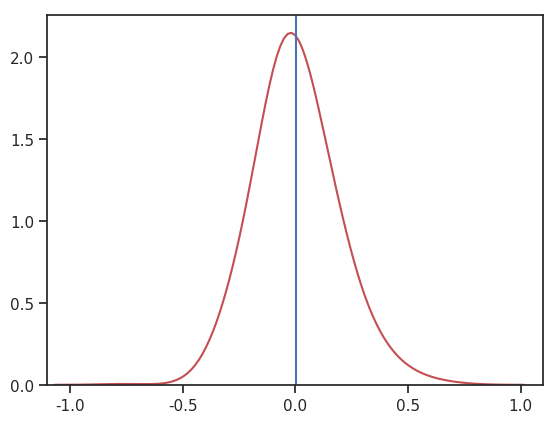}}
        % labels
        \put(-1,9){\rotatebox{90}{\footnotesize Density}}
        \put(-1,38){\rotatebox{90}{\footnotesize Density}}
        \put(34,9){\rotatebox{90}{\footnotesize Density}}
        \put(34,38){\rotatebox{90}{\footnotesize Density}}
        \put(69,9){\rotatebox{90}{\footnotesize Density}}
        \put(69,38){\rotatebox{90}{\footnotesize Density}}
        \put(15,27){\footnotesize $w_{0,0,1,1}$}
        \put(50,27){\footnotesize $w_{1,1,1,0}$}
        \put(84,27){\footnotesize $w_{2,1,3,3}$}
        \put(15,-2){\footnotesize $w_{3,2,2,2}$}
        \put(50,-2){\footnotesize $w_{4,1,5,7}$}
        \put(84,-2){\footnotesize $w_{5,2,12,14}$}
    \end{picture}
    \caption{Empirical posterior marginal density plots for of benchmark test III for $S=5$ with $\%1$ relative noise and using a $10\times10$ pressure sensor grid. Posterior mean of the coefficients is shown with the vertical blue line. \label{fig:}}
\end{figure}
\FloatBarrier

\begin{figure}[h] 
    \setlength{\unitlength}{0.01\textwidth} 
    \begin{picture}(105,55)
        % Top row
        \put(1,29){\includegraphics[width=0.31\textwidth]{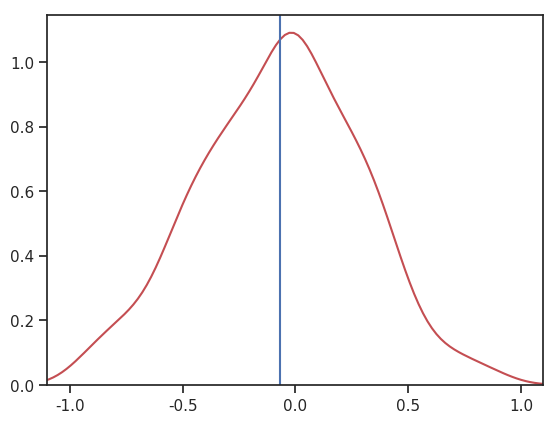}}
        \put(36,29){\includegraphics[width=0.31\textwidth]{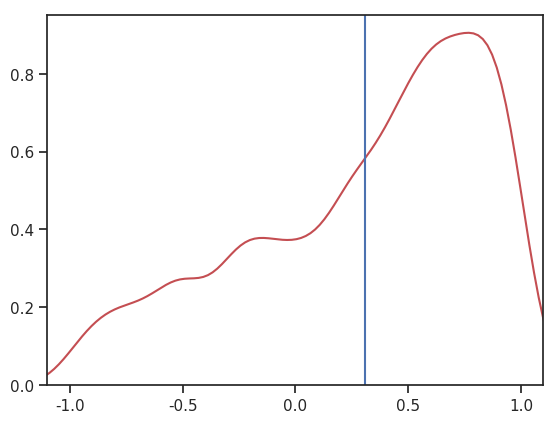}}
        \put(70.8,29){\includegraphics[width=0.31\textwidth]{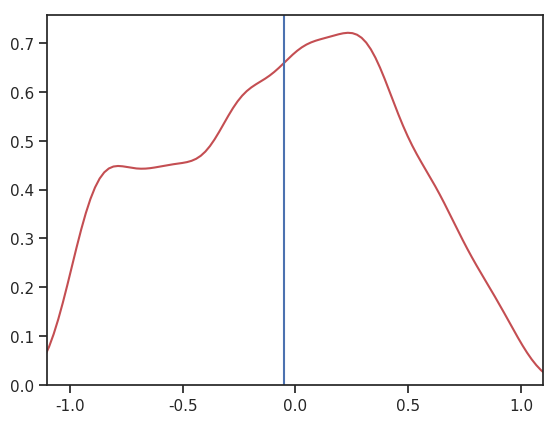}}
        % Bottom row
        \put(1,0){\includegraphics[width=0.31\textwidth]{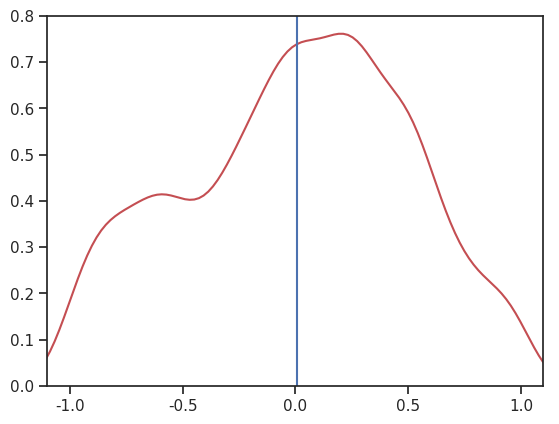}}
        \put(36,0){\includegraphics[width=0.31\textwidth]{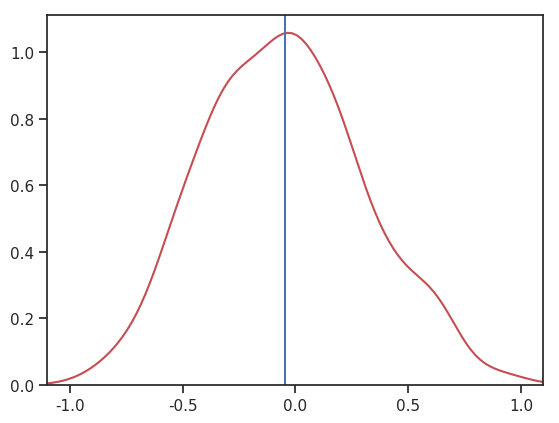}}
        \put(70.8,0){\includegraphics[width=0.31\textwidth]{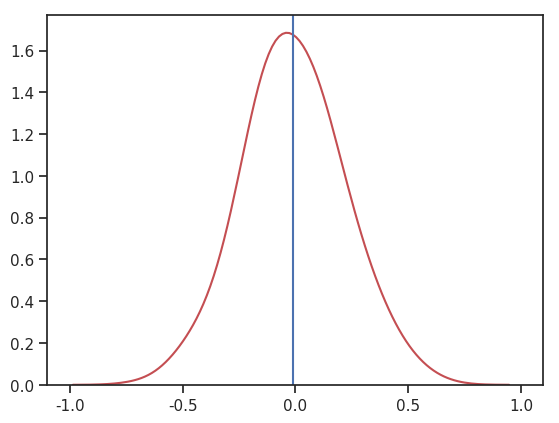}}
        % labels
        \put(-1,9){\rotatebox{90}{\footnotesize Density}}
        \put(-1,38){\rotatebox{90}{\footnotesize Density}}
        \put(34,9){\rotatebox{90}{\footnotesize Density}}
        \put(34,38){\rotatebox{90}{\footnotesize Density}}
        \put(69,9){\rotatebox{90}{\footnotesize Density}}
        \put(69,38){\rotatebox{90}{\footnotesize Density}}
        \put(15,27){\footnotesize $w_{0,0,1,1}$}
        \put(50,27){\footnotesize $w_{1,1,1,0}$}
        \put(84,27){\footnotesize $w_{2,1,3,3}$}
        \put(15,-2){\footnotesize $w_{3,2,2,2}$}
        \put(50,-2){\footnotesize $w_{4,1,5,7}$}
        \put(84,-2){\footnotesize $w_{5,2,12,14}$}
    \end{picture}
    \caption{Empirical posterior marginal density plots for of benchmark test III for $S=5$ with $\%1$ relative noise and using a $5\times5$ pressure sensor grid. Posterior mean of the coefficients is shown with the vertical blue line. \label{fig:}}
\end{figure}
\FloatBarrier

\subsection{Benchmark IV: Wrapped Gaussian process realizations}
In this benchmark test, the stochastic log-permeability field is a realization of wrapped Gaussian process, implicitly defined through a two-layer Gaussian process as follows:

\begin{equation}
    \textbf{x}'(\textbf{x})\sim \mathcal{GP}(\textbf{m}, c_1(\textbf{x}_{i,:}, \textbf{x}_{j,:}))
\end{equation}

\begin{equation}
    \text{ln } k(\textbf{x}')\sim \mathcal{GP}(0, c_2(\textbf{x}'_{i,:}, \textbf{x}'_{j,:}))
\end{equation}

\begin{equation}
    \textbf{m}(\textbf{x})=\textbf{x}
\end{equation}
where the mean function and kernels are chosen as follows
\begin{equation}
    c_1(\textbf{x}_{i,:}, \textbf{x}_{j,:}) = s_{g,1}^2\exp{-\sum_{k=1}^{d_s}(\frac{x_{ik}-x_{jk}}{l_1})^2}
\end{equation}

\begin{equation}
    c_2(\textbf{x}'_{i,:}, \textbf{x}'_{j,:}) = s_{g,2}^2\exp{-\|\frac{x'_{i,:}-x'_{j,:}}{l_2}\|}
\end{equation}

\begin{figure}[h]
	\centering
	\includegraphics[width=5.53cm]{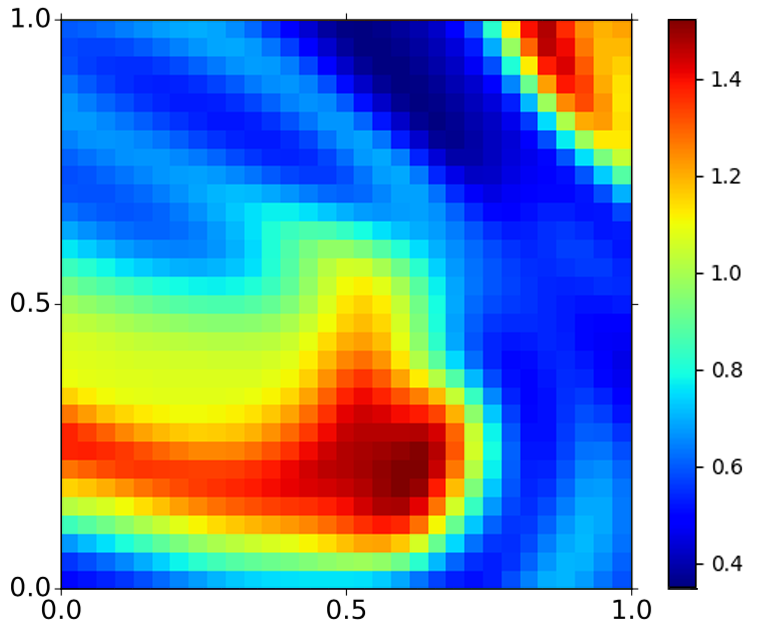}
	\caption{True log-permeability for benchmark test IV-B.}
\end{figure}

\begin{figure}[h] 
    \setlength{\unitlength}{0.01\textwidth} 
    \begin{picture}(100,67)
        % Top row
        \put(0,36){\includegraphics[width=0.31\textwidth]{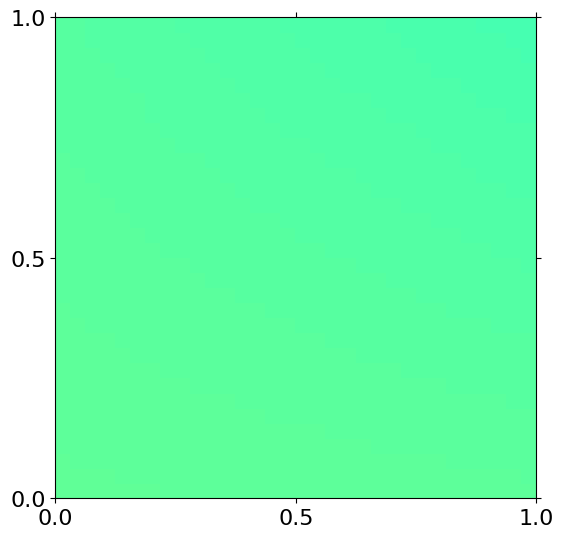}}
        \put(32,36){\includegraphics[width=0.31\textwidth]{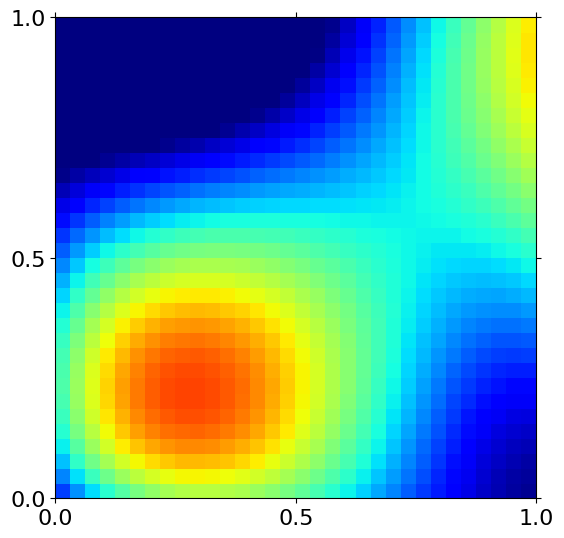}}
        \put(63.8,36){\includegraphics[width=0.31\textwidth]{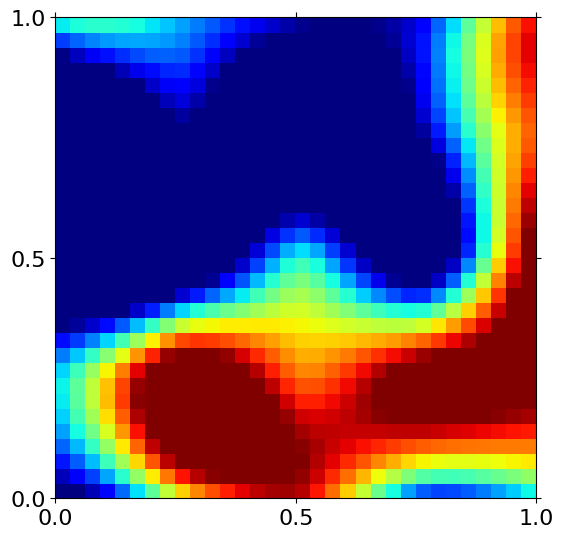}}
        % Bottom row
        \put(0,1){\includegraphics[width=0.31\textwidth]{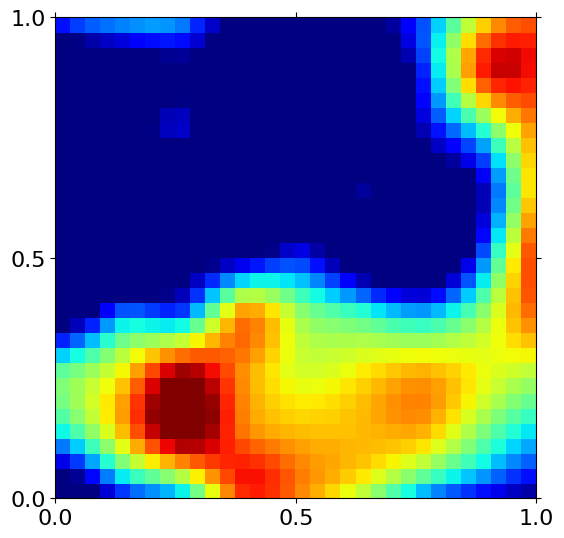}}
        \put(32,1){\includegraphics[width=0.31\textwidth]{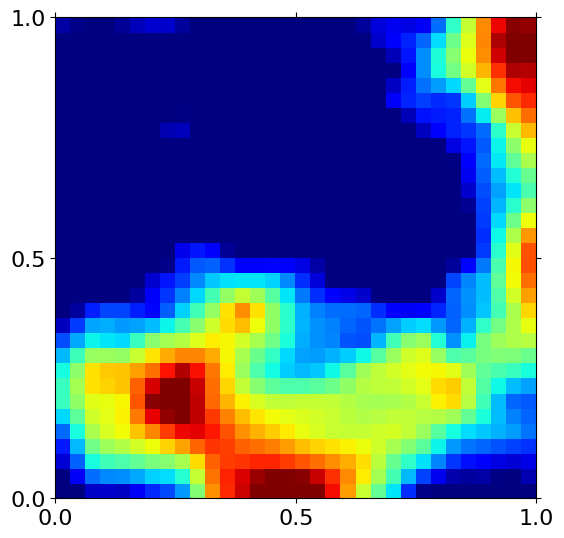}}
        \put(63.8,1){\includegraphics[width=0.31\textwidth]{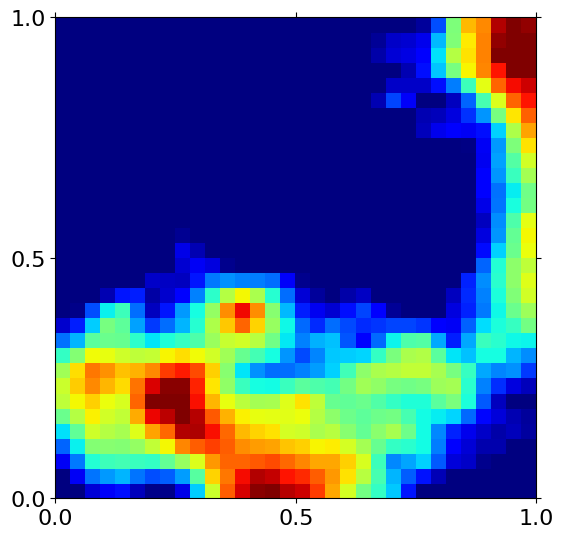}}
        % color bar
        \put(97,3){\includegraphics[width=0.0229\textwidth]{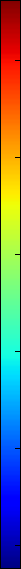}}
        \put(100,58){\footnotesize $1.4$}
        \put(100,48){\footnotesize $1.2$}
        \put(100,37){\footnotesize $1.0$}
        \put(100,26.5){\footnotesize $0.8$}
        \put(100,16){\footnotesize $0.6$}
        \put(100,5){\footnotesize $0.4$}
        % scale text
        \put(14,33){\footnotesize $S=0$}
        \put(46,33){\footnotesize $S=1$}
        \put(77.8,33){\footnotesize $S=2$}
        \put(14,-2){\footnotesize $S=3$}
        \put(46,-2){\footnotesize $S=4$}
        \put(77.8,-2){\footnotesize $S=5$}
    \end{picture}
    \caption{Posterior means of the log-permeability field of benchmark test IV-B for $S=0,\dots, 5$ with $\%5$ relative noise and using a $10\times10$ pressure sensor grid.\label{fig:BTIVC}}
\end{figure}
\FloatBarrier

\begin{figure}[h] 
    \setlength{\unitlength}{0.01\textwidth} 
    \begin{picture}(100,43)
        % Top row
        % Bottom row
        \put(17.5,-4){\includegraphics[width=0.65\textwidth]{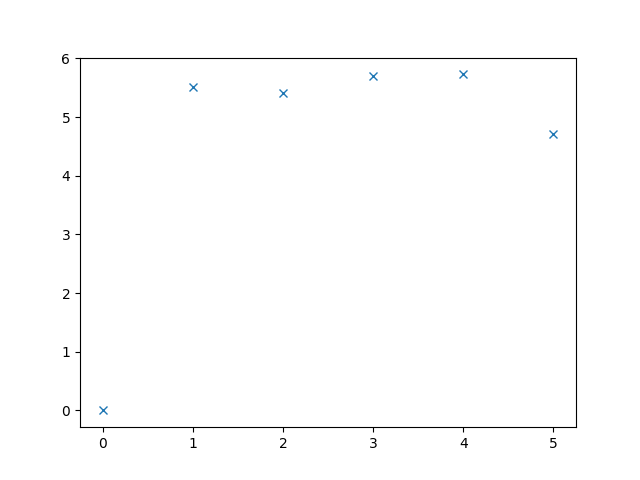}}
        % scale text
        \put(20,16){\rotatebox{90}{\footnotesize $\text{ln } BF_{S,0}$}}
        \put(50,-2.5){\footnotesize $S$}
    \end{picture}
    \caption{Log-Bayes factors for benchmark test IV-B with $\%1$ relative noise using a $10\times10$ pressure sensor network. \label{fig:BFIII}}
\end{figure}
\FloatBarrier

\begin{figure}[h]
\centering
% Subfig 1
\subfigure[]{
  \includegraphics[width=5.53cm]{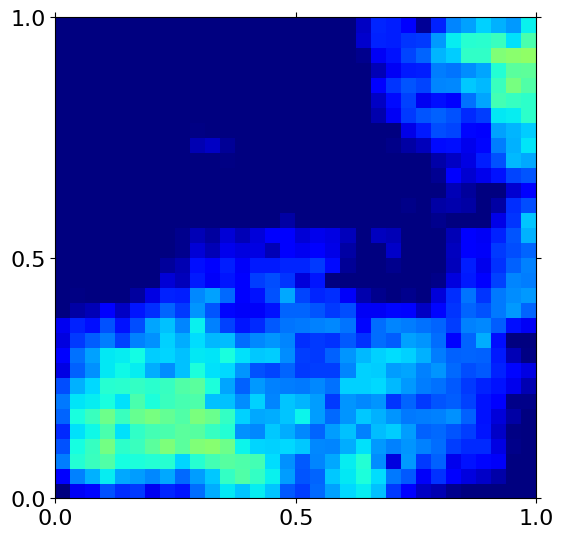}
  \label{subfig:label1}}
% Subfig 2
\subfigure[]{
  \includegraphics[width=5.53cm]{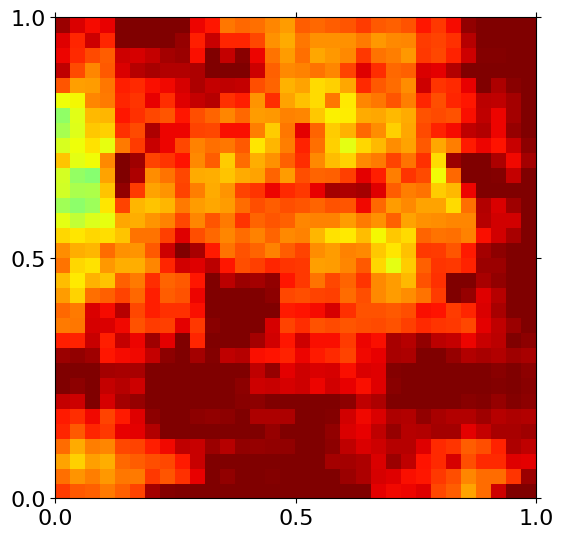}
  \label{subfig:label2}}
\caption{Posterior quantiles of the log-permeability of benchmark test IV-B for $S = 4$ using a $10 \times 10$ sensor network with $1\%$ relative noise. Left shows $5\%$ quantile and right shows $95\%$ quantile. \label{fig:errfig5}}
\end{figure}
\FloatBarrier

% \usepackage{multirow}
\begin{table}[]
    \centering
    \caption{Root mean square error for converged level of resolution and the number of non-zero bases for different benchmark tests. Here, the total number of bases at each resolution level is $2^{2(S+1)}$. \rule[-8pt]{0pt}{12pt} \label{tab:baserr}}
    \begin{tabular}{ccccccccc}
    \hline
    \multirow{2}{*}{Figure}             & \multirow{2}{*}{Test no.} & \multicolumn{6}{c}{no. of bases}              & \multirow{2}{*}{RMSE} \\ \cline{3-8}
                                        &                           & S = 0 & S = 1 & S = 2 & S = 3 & S = 4 & S = 5 &       \\ \hline
    \ref{fig:BTIC}     & I                       & 4     & 16    & 64    & 184   & 332   & 356   & $1.43 \times 10^{-1}$      \\
    \ref{fig:BTIIC}    & II                      & 4     & 16    & 64    & 220   & 732   & 2068  & $7.33 \times 10^{-1}$      \\
    \ref{fig:BTIIIC}   & III                     & 4     & 16    & 60    & 208   & 664   & 1368  & $5.72 \times 10^{-1}$   \\
    \ref{fig:BTIII5xC} & III, $5\times 5$ grid   & 4     & 16    & 64    & 212   & 636   & 1308  & $6.44 \times 10^{-1}$      \\
    \ref{fig:BTIII5pC} & III, $5\%$ noise        & 4     & 16    & 64    & 244   & 736   & 1560  & $5.12 \times 10^{-1}$     \\
    \ref{fig:BTIVC}    & IV                      & 4     & 16    & 60    & 228   & 752   & 2024  & $3.20 \times 10^{-1}$    \\ \hline
    \end{tabular}
    \vspace{12pt}
\end{table}

\begin{figure}[h]
\centering
% Subfig 1
\subfigure[$S=0$]{
  \includegraphics[width=4.7cm]{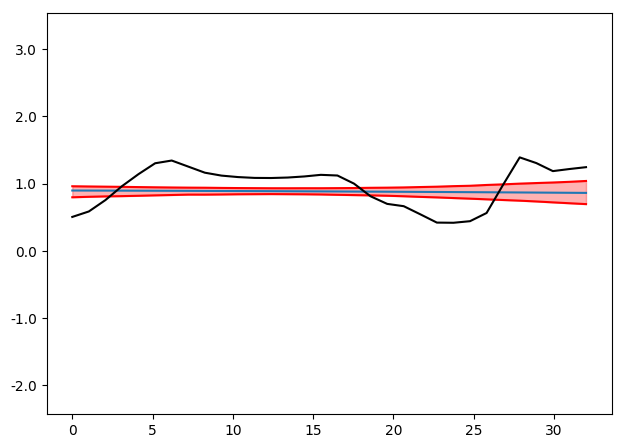}
  \label{subfig:label1}}
% Subfig 2
\subfigure[$S=1$]{
  \includegraphics[width=4.7cm]{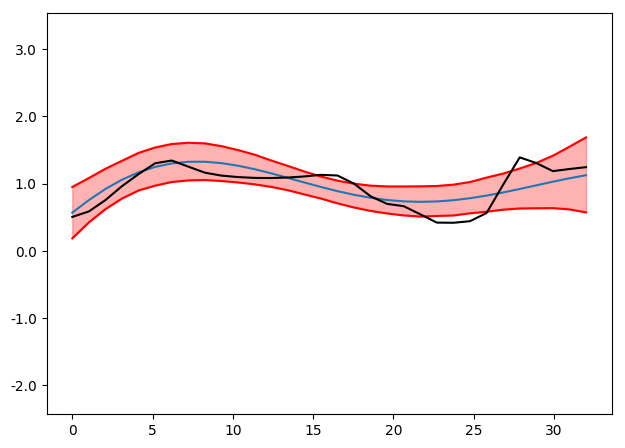}
  \label{subfig:label2}}
% Subfig 3
\subfigure[$S=2$]{
  \includegraphics[width=4.7cm]{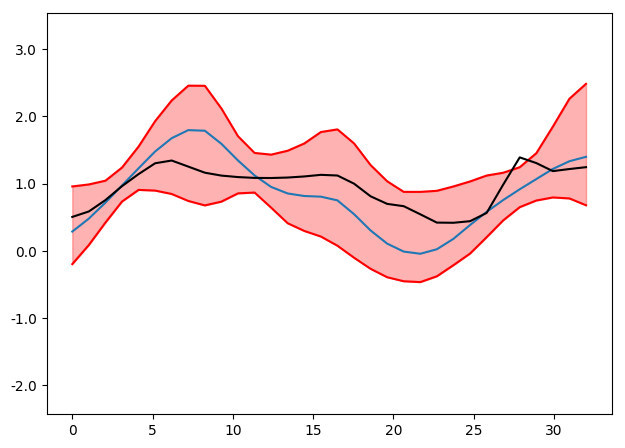}
  \label{subfig:label1}}
  % Subfig 4
\subfigure[$S=3$]{
  \includegraphics[width=4.7cm]{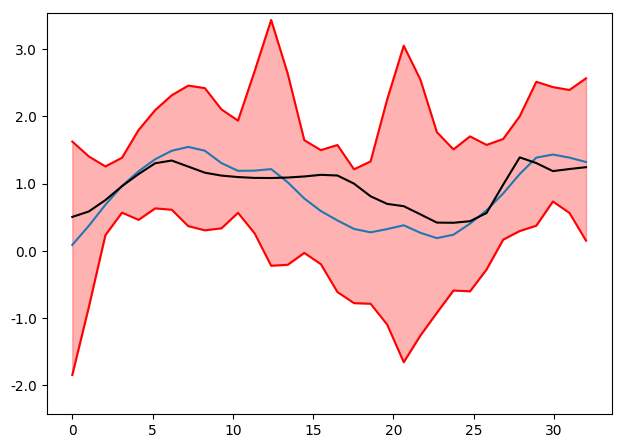}
  \label{subfig:label1}}
% Subfig 5
\subfigure[$S=4$]{
  \includegraphics[width=4.7cm]{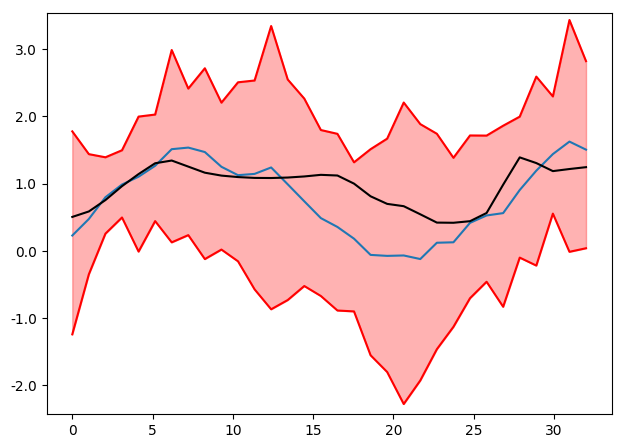}
  \label{subfig:label1}}
% Subfig 5
\subfigure[$S=5$]{
  \includegraphics[width=4.7cm]{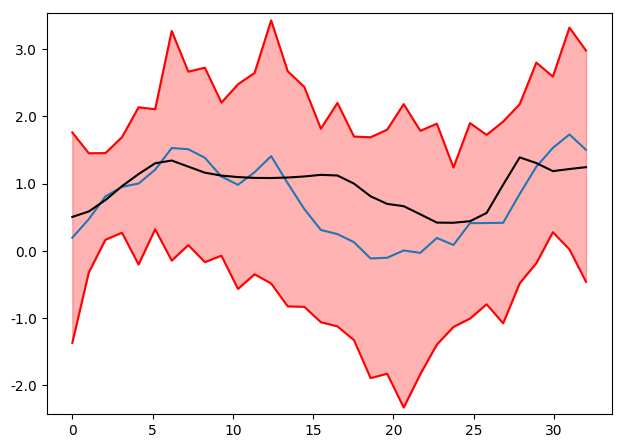}
  \label{subfig:label1}}
\caption{Posterior means of the Log-permeability field of benchmark test IV-B along the line $x=y$ for $S=0,\dots,5$ with $\%1$ noise.\label{fig:PostPIV-B}}
\end{figure}

%%% Local Variables:
%%% mode: latex
%%% TeX-master: "../JCP-CBVP"
%%% End:

% !TEX root = ../JCP-SCLS.tex
\section{Conclusions}
This work uses a multiresolution approximation to parametrize the unknown variable of a Bayesian inverse problem. This multiresolution analysis is based on second-generation wavelets which are generalization of biorthogonal wavelets and use lifting scheme and interpolating wavelet transform to define the wavelets. This choice of wavelets results in a sparser representation which effectively reduces the number of unknown parameters. Also, using lifting scheme reduces the cost of wavelet decomposition by half. A prior model was defined to incorporate belief of wavelet coefficients' quadtree structure into the inverse problem using the Bayesian methodology. This takes the sparseness and hierarchical relation between wavelet coefficients of different resolutions into account by setting a spike-and-slab distribution on the prior probability of wavelet coefficients. The effect of this choice of prior on sparseness of the model was numerically verified in table \ref{tab:baserr} using the total number of non-zero bases and comparisons were made in section \ref{Numerical results} with the results reported in \cite{ellam2016bayesian} which were achieved using a uniform prior distribution. In contrast with \cite{ellam2016bayesian} which scales the posterior of wavelet coefficients to guarantee a degree of smoothness in the final solution, here level-dependent factors are introduced in the prior model in order to define a rate of decay on the wavelet coefficients. The algorithm also takes advantage of multilevel adaptive wavelet collocation method for generating the observation data. Benchmark tests were performed to show the performance of the numerical algorithm. The result show excellent performance in finding the $90\%$ credential region, in comparison to overconfident results in \cite{ellam2016bayesian}, where the true solution is mostly located out of the bounds for all the test cases. The mean is also predicted the true solution very closely in all benchmark test cases.  Due to this choice of prior, the algorithm converges faster in comparison to algorithms with widely used priors such as a uniform distribution, and it reconstructs the field with lower number of basis and higher thresholding precision.

%%% Local Variables:
%%% mode: latex
%%% TeX-master: "../JCP-CBVP"
%%% End:
\FloatBarrier

\section{Acknowledgment\label{sec:ack}}
The author thanks Scott Hampton, Louis Ellam, and Souvik Chakraborty for their helpful discussions.

\bibliographystyle{unsrtFirstInit}
\bibliography{references}

\end{document}